\documentclass[letterpaper, 10pt]{IEEEtran}
\usepackage[top=0.75in, bottom=0.75in, left=0.625in, right=0.625in]{geometry}
\usepackage{amsmath}
\usepackage{amsthm}
\usepackage{amsfonts}
\usepackage{amssymb}
\usepackage{hyperref}
\usepackage{array}                      
\usepackage{cite}
\usepackage{mathtools}
\usepackage{comment}
\usepackage{xcolor}
\usepackage{mathrsfs}
\usepackage{caption}
\usepackage{subcaption}
\usepackage[compact]{titlesec}
\usepackage{tabularx}
\usepackage{graphicx}

\usepackage[font=small,labelfont=bf]{caption}
\usepackage [english]{babel}
\usepackage [autostyle, english = american]{csquotes}
\MakeOuterQuote{"}

\newcommand{\RR}{\mathbb{R}}
\newcommand{\ZZ}{\mathbb{Z}}
\newcommand{\dd}{{\tt d}}
\newcommand{\ball}{\mathbb{B}}

\newcommand{\D}[1]{{\tt D}(#1)}
\newcommand{\cl}[1]{\text{cl}(#1)}

\newcommand{\verts}{\mathcal{V}}

\newcommand{\nbd}{\mathcal{N}}

\newcommand{\dom}{\text{dom }}
\newcommand{\hyb}{\mathcal{H}}
\newcommand{\attr}{\mathcal{A}}
\newcommand{\statespc}{\mathcal{X}}

\newcommand{\rge}[1]{\text{rge}(#1)}

\newcommand{\wspc}{\mathcal{D}}
\newcommand{\cent}{{\tt c}}
\newcommand{\locost}{\mathcal{L}}
\newcommand{\sat}[1]{\text{sat}(#1)}

\newcommand{\tspc}{\mathcal{T}}

\theoremstyle{definition}
\newtheorem{theorem}{Theorem}

\newtheorem{definition}{Definition}

\newtheorem{lemma}{Lemma}

\newtheorem{remark}{Remark}

\newcommand{\fred}[1]{\textcolor{black}{#1}}

\begin{document}
\pagenumbering{gobble}

\title{Timer-Based Coverage Control for Mobile Sensors}
	
\author{Federico M. Zegers, Sean Phillips, and Gregory P. Hicks
\thanks{Federico M. Zegers and Gregory P. Hicks are with the Johns Hopkins University Applied Physics Laboratory, Laurel, MD 20723, USA. E-mails: {\tt \{federico.zegers,gregory.hicks\}@jhuapl.edu}.}
\thanks{Sean Phillips is with the Air Force Research Laboratory, Space Vehicles Directorate, Kirtland AFB, NM 87117, USA. 
Approved for public release; distribution is unlimited. Public Affairs
approval \#AFRL-2023-4937.
}
\thanks{This work was sponsored by the Office of Naval Research (ONR), under grant number N00014-21-1-2415. 
The views and conclusions contained herein are those of the authors only and should not be interpreted as representing those of ONR, the U.S. Navy, the U.S. Government, or any other affiliated agency.}
}

\maketitle
		
\begin{abstract} 
This work investigates the coverage control problem over a static, \fred{compact}, and convex workspace and develops a hybrid extension of the continuous-time Lloyd algorithm.
Each agent in a multi-agent system (MAS) is equipped with a timer mechanism that generates intermittent measurement and control update events, which may occur asynchronously between agents.
Between consecutive event times, as determined by the corresponding timer mechanism, the controller of each agent is held constant.
These controllers are shown to drive the configuration of the MAS into a neighborhood of the set of centroidal Voronoi configurations, i.e., \fred{the minimizers} of the standard locational cost.
The combination of continuous-time dynamics with intermittently updated control inputs is modeled as a hybrid system.
The coverage objective is posed as a set attractivity problem for hybrid systems, where an invariance-based convergence analysis yields sufficient conditions that ensure maximal solutions of the hybrid system asymptotically converge to a desired set.
A brief simulation example is included to showcase the result.
\end{abstract}

\section{Introduction}
\subsection{Motivation}
In the foundational work~\cite{Lloyd1982}, Stuart Lloyd created a variation of the method known as Pulse-Code Modulation, which uses the Shannon-Nyquist sampling theorem and signal quantization to reconstruct a band-limited signal.
According to the sampling theorem, a bounded signal $s\colon(-\infty,\infty)\to\RR$ with frequency content in a fixed interval $[0,W]$ can be recovered by utilizing
\begin{equation} \label{eqn: Shannon-Nyquist}
    s(t) = \sum_{k=-\infty}^\infty s(t_k)\frac{\sin(2\pi W(t-t_k))}{2\pi W(t-t_k)},
\end{equation}
where $s(t_k)$ is the value of $s$ at sampling time $t_k\triangleq k/(2W)$.
Rather than use a sequence of samples $\{s(t_k)\}_{k=-\infty}^\infty$ to recover $s$ with~\eqref{eqn: Shannon-Nyquist}, Lloyd's extension of Pulse-Code Modulation separates the codomain of $s$ with a centroidal Voronoi tessellation\footnote{A CVT is a Voronoi tessellation where each generator is coincident with the center of mass (centroid) of its respective Voronoi cell.
See~\eqref{eqn: Voronoi cell} and~\eqref{eqn: Centroid} for the formal definitions of a Voronoi cell and the center of mass of a Voronoi cell, respectively.} (CVT) and, for each sampling time $t_k$, sets $s(t_k)$ in~\eqref{eqn: Shannon-Nyquist} equal to the centroid of the Voronoi cell containing the sample $s(t_k)$.
Sections IV--VI in~\cite{Lloyd1982} present an algorithm for computing CVTs and their respective generators, which locally minimize a noise power cost function---the average of the squared error between the reconstructed output signal and original signal.

The result in~\cite{Lloyd1982} inspired numerous research articles whose content ranges from the rigorous analysis of continuous-time and discrete-time variants of Lloyd's algorithm, such as~\cite{Cortes.Martinez.ea2004,Du.Emelianenko2006,Du.Emelianenko.ea2006}, to applications in data compression for image processing and models of territorial behavior in animals~\cite{Du.Faber.ea1999}.
Lloyd's algorithm has also influenced the coverage control (i.e., sensor coverage) literature.
Formally, sensor coverage can be posed as an optimal sensor placement problem, which seeks to maximize a quality-of-service metric or minimize a sensor-degradation metric.
In either case, the metric is generally referred to as the \textit{locational cost}.
In~\cite{Cortes.Martinez.ea2004}, the authors developed distributed and asynchronous coverage controllers
for a multi-agent system (MAS) of mobile sensors maneuvering within a convex polytopal workspace.
The coverage controllers are based on continuous-time and discrete-time versions of Lloyd's algorithm.
LaSalle's invariance principle is used to prove all solutions of the closed-loop dynamical system, under the continuous-time Lloyd controller, asymptotically converge to the set of configurations corresponding to a CVT.\footnote{In general, the local minimizers of a locational cost are configurations that induce a CVT, which need not be unique.}
A similar argument leveraging the global convergence theorem is provided for the analogous discrete-time dynamical system. 

\subsection{Background}
Within the control theory and robotics literature, the control strategies in~\cite{Cortes.Martinez.ea2004} have been extended in diverse directions.
For instance, \cite{Santos.DiazMercado.ea2018} developed a distributed Lloyd-inspired coverage controller for agents with heterogeneous sensors, which were modeled using locational cost summands defined by distinct density functions.
The work in~\cite{Santos.Mayya.ea2019} devised a minimum-energy distributed coverage controller for a locational cost defined by a common time-dependent density function.
Unlike the closed-form controller in~\cite{Santos.DiazMercado.ea2018}, the control policy in~\cite{Santos.Mayya.ea2019} is numerically generated by solving a quadratic program with barrier function constraints pointwise in real time. 
In a similar vein, a centralized coverage controller capable of rendering the set of CVT-inducing configurations exponentially stable is extended in~\cite{Xu.DiazMercado2020} to accommodate time-varying workspace geometries.

A remarkable feature of Lloyd-based coverage controllers is their simplicity. 
Given a MAS and a bounded convex workspace tessellated by Voronoi cells defined by the instantaneous MAS configuration, the associated locational cost can be minimized simply by maneuvering each agent towards the centroid of their Voronoi cell.
Note, the centroid of an agent's Voronoi cell can be computed with knowledge of the position of the agent itself, the positions of the agent's Voronoi neighbors\footnote{Although we precisely define the notion of a Voronoi neighbor later, two agents are said to be Voronoi neighbors if their Voronoi cells share a common interface.}, the workspace, and the density function used to define the locational cost.

\subsection{Review of Prior Art}
A common thread between~\cite{Cortes.Martinez.ea2004,Santos.DiazMercado.ea2018,Santos.Mayya.ea2019,Xu.DiazMercado2020} is the reliance on continuous communication to acquire position information from Voronoi neighbors, which facilitates continuous state feedback. 
Although continuous state feedback creates robustness to small perturbations, the sustained expenditure of computational and network resources to calculate quasi-static or static Voronoi cell centroids is wasteful, especially since the centroids of a Voronoi tessellation become stationary as they approach a CVT.\footnote{The centroids of a CVT for a static workspace and static density function are stationary.
If either the workspace or density function varies with time, the centroids of a CVT may be non-stationary.}
Consequently, \cite{Nowzari.Cortes2012,Ajina.Tabatabai.ea2021,RodriguezSeda.Xu.ea2023} develop event-triggered and self-triggered Lloyd-inspired coverage controllers to more efficiently consume limited resources via intermittent communication and sensing.

In~\cite{Nowzari.Cortes2012}, the authors develop a self-triggered coverage control strategy that drives a MAS from any initial configuration to a centroidal Voronoi configuration (CVC), that is, a configuration inducing a CVT.
Each agent continuously measures their position and intermittently obtains position samples of its Voronoi neighbors via two-way communication or sensing whenever instructed by their self trigger mechanism.
The moments when a trigger mechanism causes an agent to obtain position samples from Voronoi neighbors are referred to as event times.
Between consecutive event times, an agent does not precisely know the location of their Voronoi neighbors.
To manage uncertainty, the velocity of each agent is bounded by a positive constant $v_{\max}$, which is the same for all agents.
Following the notation in~\cite{Nowzari.Cortes2012}, let $p_j^i$ and $\tau_j^i$ denote the last known position of agent $j$ stored by agent $i$ and the time elapsed since agent $i$ obtained the latest position sample from agent $j$, respectively.
Even though agent $i$ may not exactly know the position of Voronoi neighbor $j$ at every moment, the position of agent $j$ is contained within a closed ball centered about $p_j^i$ with radius $v_{\max}\tau_j^i$.
Using these closed sets, agent $i$ can compute its guaranteed Voronoi cell and dual guaranteed Voronoi cell which are under and over approximations of agent $i$'s Voronoi cell, respectively.
If each agent in the MAS moves towards the centroid of its guaranteed Voronoi cell and satisfies the inequality in~\cite[Equation 11]{Nowzari.Cortes2012} for all time, then the set of CVCs can be shown to be globally attractive.
Observe, the guaranteed and dual guaranteed Voronoi cells of agent $i$, \cite[Equation 11]{Nowzari.Cortes2012}, and~\cite[Proposition 5.2]{Nowzari.Cortes2012} are used to construct the self trigger mechanism of agent $i$ in~\cite[Table 3]{Nowzari.Cortes2012}, which allows agent $i$ to conservatively forecast when position updates from its Voronoi neighbors will be needed to satisfy~\cite[Equation 11]{Nowzari.Cortes2012} for all time.

The two-way communication strategy in~\cite{Nowzari.Cortes2012} is replaced with one-way broadcasts of positions and speed promises in~\cite{Ajina.Tabatabai.ea2021}, which utilizes event trigger mechanisms to generate intermittent broadcast events.
A speed promise, as defined in~\cite{Ajina.Tabatabai.ea2021}, consists of a positive constant, for example $s_{\max}$, and a guarantee from the broadcasting agent that it will not achieve a speed greater than $s_{\max}$ until its next broadcast event.
Therefore, if agent $j$ broadcasts a position of $p_j^i$ and a speed promise of $s_j^i$ to agent $i$, then the position of agent $j$ is contained within the closed ball centered about $p_j^i$ with radius $s_j^i\tau_j^i$.
Utilizing these closed sets, agent $i$ can compute\footnote{Note that the computation of guaranteed and dual guaranteed Voronoi cells requires information broadcasts from dual guaranteed Voronoi neighbors, i.e., Voronoi neighbors as well as potentially other agents.} its guaranteed Voronoi cell and dual guaranteed Voronoi cell, which help define the convex set in~\cite[Equation 10]{Ajina.Tabatabai.ea2021}.
Agent $i$ then projects its position onto this set and moves towards the projected point provided the distance between the projection and the position of agent $i$ is non-zero.
Otherwise, agent $i$ remains stationary and waits for information that enables motion.
Note, each agent continuously measures their position, continuously calculates the nontrivial convex set in~\cite[Equation 10]{Ajina.Tabatabai.ea2021} using sampled positions from dual guaranteed Voronoi neighbors and time, and continuously computes their projection onto the convex set.
If every agent in the MAS moves towards their projected point and satisfies~\cite[Corollary 1]{Ajina.Tabatabai.ea2021} for all time, then the set of CVCs can be shown to be globally attractive.
Since every agent must know the speed promise of all dual guaranteed Voronoi neighbors to properly compute their guaranteed and dual guaranteed Voronoi cells, each agent broadcasts information whenever they update their speed.
Consequently, the event trigger mechanism in~\cite[Equation 12]{Ajina.Tabatabai.ea2021} must be monitored continuously, unlike the self trigger mechanism in~\cite[Table 3]{Nowzari.Cortes2012}.

In~\cite{RodriguezSeda.Xu.ea2023}, the authors develop a self-triggered coverage control strategy that builds on~\cite{Nowzari.Cortes2012} and~\cite{Ajina.Tabatabai.ea2021} and renders a neighborhood of the set of CVCs locally attractive.
Like the strategy in~\cite{Nowzari.Cortes2012}, whenever the self trigger mechanism of an agent generates an event, the corresponding agent acquires position samples of all Voronoi neighbors simultaneously via two-way communication or sensing.
In addition, each agent continuously measures their position, like in~\cite{Ajina.Tabatabai.ea2021}.
Since every agent knows the instantaneous positions of itself and its Voronoi neighbors at each event time, each agent can intermittently compute their Voronoi cell and the centroid of their Voronoi cell.
Hence, each agent can maneuver itself towards the centroid of its most recently sampled Voronoi cell by utilizing the saturation controller in~\cite[Equation 5]{RodriguezSeda.Xu.ea2023}.
Since the controller is saturated and employs the same positive parameters $\kappa$ and $v$, the velocity of every agent is bounded by the constant $v$.
Let the event time $t_k^i$ denote the $k^{\text{th}}$ instant agent $i$ obtains position information from all its Voronoi neighbors.
As a result of the saturation controller, for each agent $i$, the position of Voronoi neighbor $j$ is contained within the closed ball centered about $p_j(t_k^i)$ with radius $v\cdot(t-t_k^i)$.
Note, $p_j(t_k^i)$ represents the position of agent $j$ sampled at event time $t_k^i$ and $t\in[t_k^i,t_{k+1}^i)$ denotes the instantaneous time.
With these closed sets, each agent can compute their guaranteed Voronoi cell and dual guaranteed Voronoi cell.
Note, the self trigger mechanism of agent $i$ in~\cite[Equation 8]{RodriguezSeda.Xu.ea2023} employs the guaranteed and dual guaranteed Voronoi cells of agent $i$ to ensure two properties.
One, for every agent $i$ and all time, the distance between the sampled and instantaneous Voronoi cell centroids of agent $i$ is upper bounded by a positive parameter $\delta$.
Two, for every agent $i$ and each flow interval, the sampled Voronoi cell centroid of agent $i$ is contained within the guaranteed Voronoi cell of agent $i$.
Observe, the second property guarantees the centroids of any two adjacent Voronoi cells have strictly positive distance during flows since guaranteed Voronoi cells are disjoint.
In addition, the self trigger mechanism of~\cite{RodriguezSeda.Xu.ea2023} is meticulously designed to exclude Zeno behavior, which is not considered in~\cite{Nowzari.Cortes2012} and~\cite{Ajina.Tabatabai.ea2021}.
In particular, if the distance between any pair of agents can be lower bounded by a small positive parameter $\epsilon$ at each event time and the distance between any agent and the boundary of the workspace can be lower bounded by $\epsilon/2$ at each event time, then there exists a positive constant $\tau^{\star}$ that lower bounds the distance between any pair of consecutive event times for each agent.
Furthermore, given a desired minimum inter-event time $\tau_{\min}\in\RR_{>0}$, \cite[Corollary 1]{RodriguezSeda.Xu.ea2023} provides sufficient conditions for ensuring $\tau^{\star}\geq\tau_{\min}$.

\subsection{Contributions}
Motivated by the results in~\cite{Nowzari.Cortes2012,Ajina.Tabatabai.ea2021,RodriguezSeda.Xu.ea2023}, we develop a distributed timer-based coverage controller that renders a neighborhood of the set of CVCs locally attractive.
Rather than employ event or self trigger mechanisms to create communication/measurement and control update events, such as in~\cite{Nowzari.Cortes2012,Ajina.Tabatabai.ea2021,RodriguezSeda.Xu.ea2023}, we utilize timer mechanisms instead. 
The use of timer mechanisms is prompted by our previous contributions to the consensus literature, for example, \cite{Phillips.Sanfelice2019,Zegers.Phillips.ea2021,Nino.Zegers.ea2023}, which employ timer-based control to produce intermittent sampling events which may occur asynchronously between agents.
Within this work, each agent is furnished with their own timer mechanism, and, when the timer mechanism of an agent generates an event, two ordered actions take place: 1) the corresponding agent measures the positions of itself and its Voronoi neighbors and 2) the control input of the corresponding agent is updated using the new information.
To be more precise, at each event time, agent $p$ calculates the centroid of its Voronoi cell based on the position samples and maneuvers towards the sampled centroid by employing a hybrid controller.
Therefore, every agent acquires position information and performs control updates in an intermittent manner.
Moreover, since each agent has a personal timer mechanism that can generate event times independent of other timers, the event times of any two distinct agents may be asynchronous.

The proposed timers provide several additional benefits.
For instance, using their timer mechanism, each agent can precisely determine their next event time at each event time, much like a self trigger mechanism.
Hence, there is no need to continuously monitor a trigger mechanism, unlike~\cite{Ajina.Tabatabai.ea2021}. 
Further, the proposed timer mechanisms do not require the use of guaranteed and/or dual guaranteed Voronoi cells, which provides a computational savings not enjoyed by the triggering strategies in~\cite{Nowzari.Cortes2012,Ajina.Tabatabai.ea2021,RodriguezSeda.Xu.ea2023}. 
Unlike the event/self trigger mechanisms of~\cite{Nowzari.Cortes2012,Ajina.Tabatabai.ea2021,RodriguezSeda.Xu.ea2023}, our timer mechanisms enable the construction of \textit{all} event-time sequences where the distance between any consecutive pair of event times is bounded above and below by user-defined constants.\footnote{The timer mechanisms can be used to model packet dropouts and missed resets, as discussed in Remark~\ref{rem: timer discussion}.}
Thus, Zeno behavior in the timers is excluded \textit{a priori}, and the lower bounding constant can be selected to respect the sampling rate of the cyber-physical system at hand.
Using a similar strategy to that in~\cite{Zegers.Deptula.ea2023}, which studies the rendezvous problem, we develop a maximum dwell-time condition that prescribes the size of the upper bounding constant to ensure the convergence result.

Each agent is modeled as a single-integrator in $n$-dimensional Euclidean space and employs a hybrid controller to achieve the coverage objective.
Consequently, the ensemble is modeled as a hybrid system, which is analyzed using the framework of~\cite{Goebel.Sanfelice.ea2012}.
In our analysis, we demonstrate all maximal solutions of the closed-loop hybrid system are complete and non-Zeno, which facilitates the study of the asymptotic behavior of solutions---such a consideration is not discussed in~\cite{Nowzari.Cortes2012} or~\cite{Ajina.Tabatabai.ea2021}.
Moreover, an invariance-based convergence analysis for hybrid systems guides the derivation of sufficient conditions whose satisfaction guarantees all maximal solutions of the proposed hybrid system asymptotically approach a neighborhood of the set of CVCs, which we denote by $\attr$.
Note that the set $\attr$ can be adjusted via a positive user-defined parameter $\nu$, which controls the maximum distance an agent can be from its Voronoi cell centroid at steady state.
Hence, $\nu$ can be viewed as a performance level, where a small $\nu$ value is preferable.
Given a particular choice for $\nu$, the maximum dwell-time conditions, i.e., the sufficient conditions for the invariance result, provide a tool for identifying sufficient measurement rate specifications that enable $\attr$ to be an attractor.
With respect to the coverage control strategies in~\cite{Cortes.Martinez.ea2004}, \cite{Nowzari.Cortes2012,Ajina.Tabatabai.ea2021,RodriguezSeda.Xu.ea2023}, our coverage control result contributes
\begin{enumerate}
    \item a nominally well-posed hybrid system with complete and non-Zeno solutions,
    \item a sufficient condition for attractivity that relates coverage performance to measurement rate requirements, and 
    \item a control strategy that is computationally lighter, simpler to implement, and capable of achieving coverage performance commensurate with~\cite{Cortes.Martinez.ea2004} and~\cite{RodriguezSeda.Xu.ea2023} while also requiring less frequent position measurements as demonstrated in simulation.
\end{enumerate}
A $30$-agent numerical example is used to validate the theoretical development.
Furthermore, using a $12$-agent MAS, we compare our timer-based coverage controller, the continuous-time Lloyd algorithm from~\cite{Cortes.Martinez.ea2004}, and the self-triggered coverage controller from~\cite{RodriguezSeda.Xu.ea2023} via simulation.

\section{Preliminaries} \label{sec: Preliminaries Section}
\subsection{Notation}
Let $\RR$ and $\ZZ$ denote the set of reals and integers, respectively. 
For $a\in\RR$, let $\RR_{\geq a}\triangleq[a,\infty)$,
$\RR_{>a}\triangleq(a,\infty)$, $\ZZ_{\geq a}\triangleq\RR_{\geq a}\cap\ZZ$, and $\ZZ_{>a}\triangleq\RR_{>a}\cap\ZZ$.
For $p,q\in\ZZ_{>0}$, the $p\times q$
zero matrix and the $p\times 1$ zero column vector are respectively denoted by
$0_{p\times q}$ and $0_p$.
The $p\times p$ identity
matrix and the $p\times 1$ column vector with all entries being one are respectively denoted by $I_p$
and $1_p$. 
The Euclidean norm of $r\in\RR^p$
is denoted by $\Vert r\Vert \triangleq\sqrt{r^\top r}$.
For $x,y\in\RR^p$, the inner product between $x$ and $y$
is denoted by $\langle x,y\rangle\triangleq x^\top y$. 
For $M\in\ZZ_{\geq 2}$, let $[M]\triangleq\{1,2,...,M\}$.
%
%
For a collection of vectors $\{z_1,z_2,...,z_p\}\subset\RR^q$, let $(z_k)_{k\in[p]}\triangleq [z_1^\top,z_2^\top,...,z_p^\top]^\top\in\RR^{pq}$.
Similarly, for $x\in\RR^p$ and $y\in\RR^q$, let $(x,y)\triangleq [x^\top,y^\top]^\top\in\RR^{p+q}$.
For $x\in\RR^n$ and $c\in\RR_{>0}$, let
\begin{equation*}
    \sat{x,c} \triangleq \left\{
    \begin{array}{lc}
        x/c, & \Vert x \Vert \leq c\phantom{,} \\
        x/\Vert x \Vert, & \Vert x \Vert > c.
    \end{array} \right.
\end{equation*}
For $x\in\RR^q$ and $S\subset\RR^q$, let $x+S\triangleq \{x+s\in\RR^q\colon s\in S\}$.
\fred{Let $\ball^n \triangleq \{x\in\RR^n\colon \Vert x \Vert \leq 1 \}$ denote the closed unit $n$-ball centered about the origin.}
The boundary and closure of a set $Q\subset\RR^n$ are represented by $\partial Q$ and $\cl{Q}$, respectively.
For any sets $A$ and $B$, a function $f$ of $A$ with values in $B$ is denoted by $f\colon A\to B$, whereas $f\colon A\rightrightarrows B$ refers to a set-valued function $f\colon A\to 2^B$.
For $m,n\in\ZZ_{\geq 1}$, the Jacobian of a differentiable function $f\colon\RR^n\to\RR^m$ at the point $s\in\RR^n$ is denoted by $\D{f(s)}\in\RR^{m\times n}$.

\subsection{Hybrid Systems} \label{sec: hybrid systems}
Within this work, we will employ the hybrid systems framework described in~\cite{Goebel.Sanfelice.ea2012}, where the following is provided for convenience.
For a general hybrid system $\hyb$ with data $(C,F,D,G)$ and a continuously differentiable function $V\colon\RR^n\to\RR$, let
\begin{equation*}
\begin{aligned}
    u_C(z) &\triangleq \left\{
        \begin{array}{lc}
        \max_{v\in F(z)} \langle \nabla V(z), v\rangle, & z\in C\\
        -\infty, & \text{otherwise},
        \end{array}
    \right. \\[5pt]
    u_D(z) &\triangleq \left\{
        \begin{array}{lc}
        \max_{g\in G(z)} \{V(g) - V(z)\}, & z\in D\\
        -\infty, & \text{otherwise}.
        \end{array}
    \right.
\end{aligned}    
\end{equation*}

\section{Problem Formulation}
Let there be a MAS of $N\in\ZZ_{\geq 2}$ mobile sensors, which are enumerated by the elements of $\verts\triangleq [N]$.
The kinematic model of agent $p\in\verts$ is
\begin{equation}\label{eqn: agent kinematics}
    \dot{x}_p = u_p,
\end{equation}
where $x_p,u_p\in\RR^n$ denote the position and control input of agent $p$, respectively.
The configuration of the MAS, that is, the spatial arrangement of all agents, is represented by the variable $x\triangleq(x_p)_{p\in\verts}\in\RR^{nN}$.
Moreover, let there be a static, \fred{compact}, and convex workspace $\wspc\subset\RR^n$, where the relative importance between the points in $\wspc$ is determined by a user-defined and continuous density function $\varphi\colon\wspc\to\RR_{> 0}$.

The objective of this work is to develop a distributed controller for each agent $p\in\verts$ that enables the MAS to achieve sensor coverage of the workspace $\wspc$.
Each controller should use intermittent information and facilitate asynchronous operation between agents to efficiently leverage network resources and readily integrate in hardware. 

To decompose the sensor coverage objective into agent-level subtasks, consider a Voronoi tessellation of the workspace $\wspc$, where 
\begin{equation}\label{eqn: Voronoi cell}
    V_p(x) \triangleq \{z\in\wspc\colon \forall_{r\in\verts\setminus\{p\}} \ \Vert x_p - z \Vert \leq \Vert x_r - z \Vert \}
\end{equation}
denotes the Voronoi cell of agent $p$ under an ensemble configuration $x\in\wspc^N$.
The Voronoi cell $V_p(x)$ represents the region of the workspace $\wspc$ that agent $p$ is responsible for monitoring.
Sensor coverage, as discussed in~\cite{Cortes.Martinez.ea2004}, can be posed as an optimal sensor placement problem, where optimality is typically investigated relative to the locational cost $\locost\colon\wspc^N\to\RR_{> 0}$,
\begin{equation}\label{eqn: Locational Cost}
    \locost(x) \triangleq \sum_{p\in\verts} \int_{V_p(x)} \Vert x_p - z \Vert^2\varphi(z)\dd z.
\end{equation}
Let $\cent_p\colon\wspc^N\to\RR^n$ denote the centroid of the Voronoi cell of agent $p$ relative to the density function $\varphi$, such that
\begin{equation}\label{eqn: Centroid}
    \cent_p(s) \triangleq \frac{\int_{V_p(s)}z\varphi(z)\dd z}{\int_{V_p(s)}\varphi(z)\dd z}.
\end{equation}
Similarly, let $m_p\colon\wspc^N\to\RR_{> 0}$ denote the mass of the Voronoi cell of agent $p$ relative to the density function $\varphi$, such that
\begin{equation}\label{eqn: mass}
    m_p(s) \triangleq \int_{V_p(s)}\varphi(z)\dd z.
\end{equation}
Observe that $\cent_p(s)$ and $m_p(s)$ are continuous functions of the configuration $s\in\wspc^N$.

Although agent $p$ can be positioned anywhere within $V_p(x)$ to facilitate monitoring, the centroid $\cent_p(x)$ is a fitting location since the relative importance between all points in $V_p(x)$ is balanced precisely at $\cent_p(x)$.
Therefore, let the centroid tracking error of agent $p$ be given by\footnote{For each $p\in\verts$, the centroid tracking error $e_p$ is a function of $x$ as denoted by $e_p(x)$ in~\eqref{eqn: Tracking error}.
However, we write $e_p$ instead of $e_p(x)$ for notational brevity.}
\begin{equation}\label{eqn: Tracking error}
    e_p(x) \triangleq \cent_p(x) - x_p \in\RR^n.
\end{equation}
In addition, the centroid tracking error of the mobile sensor ensemble is denoted by $e\triangleq(e_p)_{p\in\verts}\in\RR^{nN}$.
Whenever $e_p=0_n$ for every $p\in\verts$, the corresponding MAS configuration induces a CVT, that is, a Voronoi tessellation where each agent position is coincident with the centroid of its respective Voronoi cell.
CVTs are not necessarily unique~\cite{Cortes.Martinez.ea2004}.
However, configurations that produce a CVT correspond to local minimizers of the locational cost defined in~\eqref{eqn: Locational Cost}.
Consequently, the MAS is said to accomplish sensor coverage whenever $e_p=0_n$ for all $p\in\verts$.
Nevertheless, given a small user-defined constant $\nu$, achieving $\Vert e_p \Vert \leq \nu$ for all $p\in\verts$ is often sufficient in practice, which motivates the following definition.
\begin{definition}
Let $\nu\in\RR_{>0}$ be a user-defined constant.
A MAS with state vector $x$ is said to achieve \textit{$\nu$-approximate coverage} of a static, \fred{compact}, and convex workspace $\wspc$ with density $\varphi$ whenever $\Vert e_p(x) \Vert \leq \nu$ for all $p\in\verts$.
\hfill$\triangle$
\end{definition}

\section{Hybrid System Modeling}
Let $0<T_1^p\leq T_2^p$ be user-defined parameters for each agent $p\in\verts$.
Moreover, let $\tau_p\in[0,T_2^p]$ be a timer variable for agent $p$, which evolves according to the hybrid system
\begin{equation}\label{eqn: Timer}
\begin{aligned}
    \dot{\tau}_p &= -1, & \tau_p &\in [0,T_2^p]\\
    \tau_p^+ &\in [T_1^p,T_2^p], & \tau_p &= 0.
\end{aligned}
\end{equation}
Note that the initial condition is given by $\tau_p(0,0)\in[T_1^p,T_2^p]$.
The hybrid system in~\eqref{eqn: Timer} enables the construction of increasing sequences of time, e.g., $\{t_k^p\}_{k=0}^\infty$, given a complete solution $\phi_\tau$, where the event time $t_k^p$ represents the $k^\text{th}$ instant $\tau_p=0$.
In addition, under~\eqref{eqn: Timer} and for all $k\in\ZZ_{\geq 0}$, the difference between consecutive event times can be bounded as
\begin{equation*}
    T_1^p \leq t_{k+1}^p - t_k^p \leq T_2^p.
\end{equation*}
\begin{remark} \label{rem: timer discussion}
    The timer mechanism in~\eqref{eqn: Timer} with initial condition $\tau_p(0,0)\in[T_1^p,T_2^p]$ can generate every sequence of event times where the distance between any two consecutive event times is bounded below and above by $T_1^p$ and $T_2^p$, respectively.
    Further, the value of $\tau_p$ after each reset can be set to any number within the interval $[T_1^p,T_2^p]$ using a deterministic and/or probabilistic rule.
    Hence, $\tau_p$ allows for the abstraction and consideration of packet dropouts and missed resets.
    For example, if a system is designed to obtain feedback whenever $\tau_p=0$ and $\tau_p$ evolves according to~\eqref{eqn: Timer}, then $T_2^p$ represents the longest amount of time the system will operate without feedback and must be selected to ensure a desired level of performance.
    \hfill$\triangle$
\end{remark}
Let $k_1\in\RR_{>0}$ and $\varepsilon\in(0,1)$ be user-defined parameters. 
Also, let $\tilde{\nu}\triangleq (1-\varepsilon)\nu$ and $\eta_p\in\RR^n$ be an auxiliary parameter and variable, respectively.
The controller of agent $p$ is designed as $u_p\triangleq \eta_p$, where $\eta_p$ evolves according to the hybrid system
\begin{equation}\label{eqn: eta_p}
\begin{aligned}
    \dot{\eta}_p &= 0_n, & \tau_p &\in [0,T_2^p]\\
    \eta_p^+ &= k_1\sat{e_p,\tilde{\nu}}, & \tau_p &= 0.
\end{aligned}
\end{equation}
Observe that, under~\eqref{eqn: Timer} and~\eqref{eqn: eta_p}, the event times in the sequence $\{t_k^p\}_{k=0}^\infty$ coincide with the instants agent $p$ samples its centroid tracking error $e_p$, which is used to update $\eta_p$ according to the jump equation in~\eqref{eqn: eta_p}.
The combination of $u_p=\eta_p$ and~\eqref{eqn: eta_p} gives rise to a sample-and-hold controller with aperiodic event times (that is, sampling) provided $0 < T_1^p < T_2^p$.\footnote{If $0<T_1^p=T_2^p$, then a periodic sample-and-hold controller is obtained.} 
To facilitate the subsequent stability analysis, let $\eta\triangleq(\eta_p)_{p\in\verts}\in\RR^{nN}$ and $\tau\triangleq(\tau_p)_{p\in\verts}\in\RR^N$ denote the control input and timer variable of the ensemble, respectively.
Moreover, let 
\begin{equation}\label{eqn: eta_p tilde}
    \tilde{\eta}_p \triangleq \eta_p - k_1\sat{e_p,\tilde{\nu}} \in\RR^n
\end{equation}
denote the sample-and-hold error of agent $p\in\verts$.

When $\tau_p=0$ for any agent $p\in\verts$, a jump may occur.
If a jump does occur, the jump equation in~\eqref{eqn: Timer} implies that $\tau_p$ is reset to some value within the interval $[T_1^p,T_2^p]$.
Additionally, the agent kinematics in~\eqref{eqn: agent kinematics} indicate that $x_p^+=x_p$, after any jump, for all $p\in\verts$.
Hence, $x^+=x$.
Moreover, the continuous dependence of $e_p$ on $x$, as defined in~\eqref{eqn: Tracking error}, implies that $e_p^+=e_p$ after any jump.
However, the jump equation in~\eqref{eqn: eta_p} and the definition in~\eqref{eqn: eta_p tilde} indicate that $\eta_p^+=k_1\sat{e_p,\tilde{\nu}}$ and $\tilde{\eta}_p^+ = 0_n$ after a jump triggered by $\tau_p=0$.
If a jump is only caused by $\tau_q=0$ for some $q\in\verts$, then, for $p\neq q$, $\eta_p^+=\eta_p$, $\tilde{\eta}_p^+=\tilde{\eta}_p$, and $\tau_p^+=\tau_p$.
In other words, $\eta_p$, $\tilde{\eta}_p$, and $\tau_p$ are mapped to themselves if and only if $\tau_p$ does not trigger a jump.

The hybrid system of the mobile sensor ensemble is denoted by $\hyb$.
The state variable and state space of $\hyb$ are denoted by $\xi\triangleq (x,\eta,\tau)\in\statespc$ and $\statespc\triangleq \RR^{nN}\times\RR^{nN}\times\RR^N$, respectively.
Furthermore, the flow and jump sets of $\hyb$ are
\begin{equation}\label{eqn: Flow and Jump sets}
    C\triangleq\bigcap_{p\in\verts}\{\xi\in\statespc\colon \tau_p\in [0,T_2^p]\},\
    D\triangleq\bigcup_{p\in\verts} \{\xi\in\statespc\colon \tau_p=0\},
\end{equation}
respectively, which are closed by construction.
For each agent $p\in\verts$, let $D_p\triangleq\{\xi\in\statespc\colon \tau_p=0\}$.
The differential equation that governs the flows of $\hyb$ is $\dot{\xi}=f(\xi)$, such that the single-valued flow map $f\colon\statespc\to\statespc$, 
\begin{equation}\label{eqn: Flow map}
    f(\xi)\triangleq 
    \big(\eta, 0_{nN}, -1_N\big)
\end{equation}
is derived by substituting~\eqref{eqn: agent kinematics}, $u_p=\eta_p$, the flow equation in~\eqref{eqn: Timer}, and the flow equation in~\eqref{eqn: eta_p} into the time derivative of $\xi$.
The difference inclusion that governs the jumps of $\hyb$ is $\xi^+\in G(\xi)$, where the set-valued jump map $G\colon\statespc\rightrightarrows\statespc$ is
\begin{equation}\label{eqn: Jump map}
\begin{aligned}
    G(\xi) &\triangleq \{G_p(\xi)\colon \xi\in D_p \text{ for some }p\in\verts \}, \\
    G_p(\xi) &\triangleq 
    \begin{bmatrix}
        x \\
        [\eta_1^\top,...,\eta_{p-1}^\top,k_1\sat{e_p,\tilde{\nu}}^\top,\eta_{p+1}^\top,...,\eta_N^\top]^\top \\
        [\tau_1,...,\tau_{p-1},[T_1^p,T_2^p],\tau_{p+1},...,\tau_N]^\top
    \end{bmatrix}.
\end{aligned}
\end{equation}
Note that the jump map is obtained by using the observations presented immediately below~\eqref{eqn: eta_p tilde}.

The solutions of the hybrid system $\hyb$ with data $(C,f,D,G)$ describe the behavior of the mobile sensor ensemble.
Therefore, the MAS can be shown to accomplish $\nu$-approximate coverage of the workspace $\wspc$ by demonstrating that the set
\begin{equation*}
    \attr\triangleq\{\xi\in\fred{\wspc^N \times (k_1\ball^n)^N\times\tspc}\colon \forall_{p\in\verts} \ \Vert e_p \Vert \leq \nu \} 
\end{equation*}
is an attractor for $\hyb$\fred{, where $\tspc\triangleq [0,T_2^1]\times[0,T_2^2]\times...\times [0,T_2^N]$}.
In fact, we will \fred{demonstrate} that every maximal solution $\phi$ of $\hyb$ converges to $\attr$, which is compact,\footnote{\fred{For $\xi\in\attr$ and each $p\in\verts$, $\eta_p\in k_1\ball^n$ and $\tau_p\in [0,T_2^p]$. 
Furthermore, the preimage of a closed set under a continuous function is closed.
Hence, the configurations $x$ are contained in a closed subset of $\wspc^N$, which is compact.}} provided certain sufficient conditions are satisfied. 

The flow set $C$ and jump set $D$ are closed by construction.
In addition, the flow map $f$ is continuous, and the jump map $G$ is outer semi-continuous and locally bounded.
As a result, $\hyb$ satisfies the hybrid basic conditions~\cite[Assumption 6.5]{Goebel.Sanfelice.ea2012}, which implies $\hyb$ is nominally well-posed~\cite[Theorem 6.8]{Goebel.Sanfelice.ea2012}. 
%
%

\section{Completeness and Dwell-Time Analysis}
In preparation for the invariance-based convergence analysis of Section~\ref{sec: Convergence analysis}, two supporting results are presented next.
The following lemma shows that each maximal solution $\phi$ of $\hyb$ has an unbounded hybrid time domain, facilitating the analysis of $\phi$'s limiting behavior.
\begin{lemma}\label{lemma: Completeness}
    Every maximal solution $\phi$ of the hybrid system $\hyb$ with data $(C,f,D,G)$ is complete and non-Zeno.
    \hfill$\triangle$
\end{lemma}
\begin{proof}
    The proof leverages~\cite[Proposition 2.10]{Goebel.Sanfelice.ea2012}.
    Let $\xi^q = (x^q,\eta^q,\tau^q)$ for $q\in\{1,2\}$, with $q$ being an index and not a power.
    Suppose $\xi^1, \xi^2\in\statespc$.
    Utilizing the flow map $f$ in~\eqref{eqn: Flow map}, it follows that $f(\xi^1) - f(\xi^2)  =  Z(\xi^1 - \xi^2)$, where $Z$ is a $3\times 3$ block matrix with a norm of one. 
    Hence, $f$ is globally Lipschitz over $\statespc$.
    Consider the differential equation $\dot{\xi}=f(\xi)$ with initial condition $\phi(0,0)\in C\setminus D$.
    Then, there exists a nontrivial and maximal solution $\phi$ for $\hyb$ satisfying the initial condition.
    Further, under the construction of $\hyb$, one has $G(D)\subset C\cup D$ which implies that Item (c) in~\cite[Proposition 2.10]{Goebel.Sanfelice.ea2012} does not occur.
    Because the single-valued flow map $f$ is Lipschitz continuous, Item (b) in~\cite[Proposition 2.10]{Goebel.Sanfelice.ea2012} does not occur.
    Hence, $\phi$ is complete. 
    Moreover, the hybrid system $\hyb$ does not possess Zeno solutions given~\cite[Lemma 3.5]{Li.Phillips.ea2018}.
\end{proof}
Currently, the only restriction placed on the timer parameters $T_1^p$ and $T_2^p$ is that they should satisfy the inequality $0<T_1^p\leq T_2^p$.
While $T_1^p$ can be selected according to the sampling rate of agent $p$'s sensors, the selection of $T_2^p$ is unclear.
Certainly, $T_2^p$ cannot be selected arbitrarily large since control systems with insufficient feedback may become unstable.
Thus, a dwell-time analysis is motivated.

\begin{lemma}\label{lemma: eta_tilde bound}
    Let $L_p\in\RR_{>0}$ denote the Lipschitz constant of the function $x\mapsto\cent_p(x)$ in~\eqref{eqn: Centroid} and $\tilde{\eta}_{\max}\in\RR_{>0}$ be a user-defined parameter.
    Suppose $T_1^p\leq T_2^p$ for each $p\in\verts$.
    If \fred{every} $T_2^p$ is selected to satisfy the \textit{maximum dwell-time condition}
    \begin{equation}\label{eqn: Maximum dwell-time}
        T_2^p \leq \frac{\tilde{\eta}_{\max} \tilde{\nu}}{k_1^2\big(L_p\sqrt{N} + 1\big)},
    \end{equation}
    then, for each maximal solution $\phi$ of $\hyb$, the sample-and-hold error in~\eqref{eqn: eta_p tilde} is uniformly bounded, i.e., $\Vert \tilde{\eta}_p(\phi(t,j)) \Vert\leq \tilde{\eta}_{\max}$ for all $(t,j)\in\dom\phi$.\footnote{For each $p\in\verts$, the uniform bound on $\tilde{\eta}_p$ introduces a restriction on the set of initial conditions.
    Specifically, every maximal solution $\phi$ of $\hyb$ must satisfy $\phi(0,0)\in\{\xi\in\statespc\colon \forall_{p\in\verts} \ \Vert \tilde{\eta}_p\Vert\leq \tilde{\eta}_{\max}\}$.}
    \hfill$\triangle$
\end{lemma}
\begin{proof}
Select a $p\in\verts$, and let $\phi$ be a maximal solution of the hybrid system $\hyb$. 
Note that $\tilde{\eta}_p$, as defined in~\eqref{eqn: eta_p tilde}, is locally Lipschitz in $(\eta_p,x)$.
Next, consider the function $\tilde{\eta}_p\mapsto \Vert \tilde{\eta}_p \Vert$.
Since $\phi$ is absolutely continuous and both the Euclidean norm and $\tilde{\eta}_p$ are locally Lipschitz functions, the function $\Vert \tilde{\eta}_p(\phi)\Vert$ is absolutely continuous and differentiable almost everywhere in $\dom\phi$~\cite{Shevitz.Paden1994}.
Thus, the time derivative of~\eqref{eqn: eta_p tilde} is
\begin{equation} \label{eqn: etaTilde_pDot}
    \dot{\tilde{\eta}}_p = \left\{
        \begin{array}{lc}
            -\frac{k_1}{\tilde{\nu}}\dot{e}_p , & \Vert e_p \Vert \leq \tilde{\nu}\phantom{,} \\
            -\frac{k_1}{\Vert e_p \Vert}\left(I_n - \frac{e_p e_p^\top}{\Vert e_p \Vert^2} \right)\dot{e}_p, & \Vert e_p \Vert > \tilde{\nu}
            \end{array}
    \right.
\end{equation}
given the definition of the saturation function in Section~\ref{sec: Preliminaries Section} and $\dot{\eta}_p=0_n$ under the flow equation in~\eqref{eqn: eta_p}.
Since $I_n - e_p e_p^\top/\Vert e_p \Vert^2$ is a projection with unit norm, 
\begin{equation} \label{eqn: norm etaTilde_pDot Bound}
    \Vert\dot{\tilde{\eta}}_p(\phi(t,j))\Vert \leq (k_1/\tilde{\nu})\Vert \dot{e}_p(\phi(t,j))\Vert
\end{equation}
for all but countably many $(t,j)\in\dom\phi$.
The substitution of $u_p=\eta_p$, for every $p\in\verts$, into the time derivative of~\eqref{eqn: Tracking error} along the solution $\phi$ yields
\begin{equation}\label{eqn: e_p Dot 1}
    \dot{e}_p(\phi) = \D{\cent_p(s)}\big\vert_{s=x(\phi)}\eta(\phi) - \eta_p(\phi).
\end{equation}
Note that $x\mapsto\cent_p(x)$ is a continuously differentiable function \fred{restricted to} the compact domain $\wspc^N$.
Hence, the Mean Value and Extreme Value theorems imply that $\cent_p$ is Lipschitz continuous with Lipschitz constant $L_p$.
Since $\cent_p$ is differentiable, for any unit vector $v\in\RR^{nN}$ and $h\in\RR$, such that $x+hv\in\wspc^N$, the directional derivative of $\cent_p$ at the point $x$ in the direction of $v$ is
\begin{equation*}
\begin{aligned}
    &\lim_{h \to 0} \frac{\cent_p(x+hv) - \cent_p(x)}{h} = \D{\cent_p(s)}\vert_{s=x}\cdot v \iff \\
    &\lim_{h \to 0} \frac{\cent_p(x+hv) - \cent_p(x) - \D{\cent_p(s)}\vert_{s=x}\cdot hv}{h} = 0_n.
\end{aligned}
\end{equation*}
Consequently, for any $\epsilon>0$ there exists a $\delta(\epsilon)>0$, such that, if $0<h<\delta(\epsilon)$, then
\begin{equation}\label{eqn: Directional derivative inequality}
\begin{aligned}
    \epsilon &\geq \frac{\Vert\cent_p(x+hv) - \cent_p(x) - \D{\cent_p(s)}\vert_{s=x}\cdot hv \Vert}{h}\\
    & \geq \left\vert\frac{\Vert \cent_p(x+hv) - \cent_p(x) \Vert}{h} - \Vert \D{\cent_p(s)}\vert_{s=x}\cdot v\Vert \right\vert.
\end{aligned}
\end{equation}
The second inequality in~\eqref{eqn: Directional derivative inequality} in conjunction with the Lipschitz continuity of $\cent_p$ imply that $\Vert \D{\cent_p(s)}\vert_{s=x}\cdot v\Vert \leq L_p + \epsilon$.
Since the unit vector $v$ and $\epsilon$ were arbitrarily chosen, $\Vert \D{\cent_p(x)}\Vert \leq L_p$ over $\wspc^N$.

Next, observe the auxiliary variable $\eta_p$ is a sample-and-hold approximation of $k_1\sat{e_p,\tilde{\nu}}$ along the solution $\phi$.
Therefore, for any agent $p\in\verts$ and centroid tracking error $e_p(\phi\fred{(t,j)})\in\RR^n$, $\Vert \eta_p(\phi(t,j))\Vert \leq k_1$ for all $(t,j)\in\dom\phi$.
Combining $\Vert\eta\Vert$ with the individual bounds for each $\eta_p$ along $\phi$, for each $p\in\verts$, yields $\Vert \eta(\phi(t,j)) \Vert \leq k_1\sqrt{N}$ for all $(t,j)\in\dom\phi$.

Since $\Vert \tilde{\eta}_p(\phi)\Vert$ is an absolutely continuous function, the time derivative $\frac{\dd}{\dd t}\Vert \tilde{\eta}_p(\phi)\Vert$ exists almost everywhere in $\dom\phi$ and is Lebesgue integrable.
Moreover, substituting~\eqref{eqn: e_p Dot 1} into~\eqref{eqn: norm etaTilde_pDot Bound} and then using the aforementioned bounds for $\D{\cent_p(x)}$, $\eta_p$, and $\eta$ leads to
\begin{equation*}
    \frac{\dd}{\dd t}\Vert \tilde{\eta}_p(\phi(t,j))\Vert \leq \Vert\dot{\tilde{\eta}}_p(\phi(t,j))\Vert \leq \frac{k_1^2\big(L_p \sqrt{N} + 1\big)}{\tilde{\nu}}
\end{equation*}
for almost all $(t,j)\in\dom\phi$.

Let $J_p(\phi)$ denote the set of indices $j\geq 0$ such that agent $p$ experiences a jump at time $t_j$ along $\phi$.
In particular, note that $\tau_p(\phi(t_j,j-1))=0$, where the hybrid jump times $(t_j,j-1)$, with $j\in J_p(\phi)$, are denoted by $\{(t_k^p,j_k^p-1)\}_{k=0}^\infty$ in increasing order. 
Let $t_0^p\triangleq 0$.
Hence, $\tau_p(\phi(t_k^p,j_k^p-1))=0$ for each $k\geq 0$.
Furthermore, along the solution $\phi$, one has that $\vert t_{k+1}^p - t_k^p\vert\leq T_2^p$ for all $k\geq 0$.
It is also worth recalling that $\tilde{\eta}_p(\phi(t_k^p,j_k^p))=0_n$ for all $k\geq 0$, where $\tilde{\eta}_p(\phi(t_k^p,j_k^p))$ denotes the value of $\tilde{\eta}_p$ after a jump. 
Without loss of generality, suppose $t\in[t_k^p,t_{k+1}^p]$ for a fixed $k\geq 0$.
Let $\xi=\phi(t,j_k^p)$ for the rest of the argument.
The substitution of $\xi$ into $\tilde{\eta}_p$, the integrability of $\tfrac{\dd}{\dd t}\Vert\tilde{\eta}_p(\phi(s,j_k^p))\Vert$ over $[t_k^p,t_{k+1}^p]$, and~\cite[Theorem 13]{Royden1968} yield
\begin{equation}\label{eqn: etaTilde_p norm bound}
\begin{aligned}
    \Vert \tilde{\eta}_p(\xi)\Vert  &= \Vert \tilde{\eta}_p(\phi(t_k^p,j_k^p))\Vert + 
    \int_{t_k^p}^{t} \frac{\dd}{\dd t}\Vert \tilde{\eta}_p(\phi(s,j_k^p))\Vert \dd s\\
    &\leq 0 + \int_{t_k^p}^{t_{k+1}^p} \frac{k_1^2\big(L_p \sqrt{N} + 1\big)}{\tilde{\nu}} \dd s \\
    &\leq \frac{k_1^2\big(L_p \sqrt{N} + 1\big)}{\tilde{\nu}} T_2^p.
\end{aligned}
\end{equation}
The substitution of~\eqref{eqn: Maximum dwell-time} into the bottom inequality of~\eqref{eqn: etaTilde_p norm bound} yields $\Vert \tilde{\eta}_p(\xi)\Vert \leq \tilde{\eta}_{\max}$ for all $t\in[t_k^p,t_{k+1}^p]$, which leads to the desired result.
\end{proof}
Before concluding the section, some observations are made.
The maximum dwell-time condition is inherently restrictive due to conservative bounding.
In the future, one could consider real-time estimation of $\dot{\tilde{\eta}}_p$ to develop a dynamic timer variable $T_2^p$ based on~\eqref{eqn: etaTilde_p norm bound}.
Such an estimate may enable enlargements of $T_2^p$ along trajectories, and, therefore, support less frequent sensing, especially as the centroid tracking errors $\{e_p\}_{p\in\verts}$ converge to small neighborhoods of the origin.

Given a desired collection of timer constants $\{T_2^p\}_{p\in\verts}$, the maximum dwell-time condition in~\eqref{eqn: Maximum dwell-time} can be expressed and used as a maximum gain condition at the expense of additional complexity.
The derivation of~\eqref{eqn: etaTilde_p norm bound}, which enables the derivation of the \textit{maximum gain condition} 
\begin{equation} \label{eqn: Maximum gain condition}
    k_1 \leq \sqrt{\frac{\tilde{\eta}_{\max} \tilde{\nu}}{T_2^p(L_p\sqrt{N}+1)}},
\end{equation}
is based on all agents employing the same value for the constant $k_1$.
To ensure a homogeneous value for $k_1$, \eqref{eqn: Maximum gain condition} in conjunction with a gossip protocol such as that in~\cite{Lu.Tang.ea2011} can be used to find a $k_1$ that satisfies~\eqref{eqn: Maximum gain condition} for every $p\in\verts$.
If heterogeneous values for $k_1$ are desired, a different analysis is needed.

Lastly, the maximum dwell-time condition in~\eqref{eqn: Maximum dwell-time} indicates a trade-off between performance and measurement rate.
Since $\tilde{\nu}= (1-\varepsilon)\nu$ and the objective is to achieve $\nu$-approximate coverage where small values of $\nu$ are preferable, a small value for $\nu$ will force the selection of a small value for $T_2^p$, leading to more frequent measurements.
High measurement rates due to small values of $\nu$ may be assuaged by selecting small values for $k_1$; however, this will slow the MAS and potentially lead to slow convergence to $\attr$.

\section{Convergence Analysis}\label{sec: Convergence analysis}
The following objects are listed to streamline the presentation of the main result, namely, Theorem~\ref{thm: A_nu is attractive}.
For every agent $p\in\verts$, let $\nbd_p(x)\triangleq\{q\in\verts\setminus\{p\}\colon V_p(x)\cap V_q(x)\neq\varnothing\}$ represent the set of Voronoi neighbors of agent $p$ corresponding to the MAS configuration $x\in\wspc^N$.
Let 
\begin{equation} \label{eqn: agent-level locational Cost}
    \locost_p(x) \triangleq \int_{V_p(x)} \Vert x_p - z \Vert^2\varphi(z)\dd z
\end{equation}
denote the summand corresponding to agent $p$ in the locational cost provided in~\eqref{eqn: Locational Cost}.
Let $\Phi\triangleq\Phi_1\cap\Phi_2\cap\Phi_3$ denote the set of admissible initial conditions for the hybrid system $\hyb$, where\footnote{The Voronoi cell $V_p(x)$ is well-defined for each $p\in\verts$ if $x\in\wspc^N$ and no two agents are coincident.
This observation motivates our restriction on the initial condition set for $\hyb$.}
\begin{equation*}
\begin{aligned}
    \Phi_1 &\triangleq \{\xi\in\statespc\colon \forall_{p\in\verts} \ (x_p,\eta_p,\tau_p)\in (\wspc, k_1\fred{\ball^n}, [T_1^p,T_2^p])\}, \\
    \Phi_2 &\triangleq \{\xi\in\statespc\colon \forall_{p\neq q\in\verts} \ x_p \neq x_q\}\fred{,}\\
    \Phi_3 &\triangleq \{\xi\in\statespc\colon \forall_{p\in\verts} \ \Vert \tilde{\eta}_p\Vert\leq \tilde{\eta}_{\max}\}.
\end{aligned}
\end{equation*}
The invariance-based convergence analysis will be aided by the set $U\triangleq \{\xi\in\fred{\wspc^N\times(k_1\ball^n)^N\times\tspc}\colon \forall_{p\in\verts} \ \Vert e_p \Vert \geq \tilde{\eta}_{\max}\tilde{\nu}/k_1 \}$.

Let $V_{pq}(x)\triangleq\{z\in\wspc\colon\Vert x_p - z\Vert \leq \Vert x_q - z\Vert\}$ denote the set of all points in $\wspc$ that are closer to agent $p$ than agent $q$ under the MAS configuration $x$.
Using these sets, the Voronoi cell of agent $p$ can be alternatively expressed as
\begin{equation}\label{eqn: Voronoi cell alternative}
    V_p(x) = \bigcap_{q\in\verts\setminus\{p\}} V_{pq}(x).    
\end{equation}
Let $\partial V_p(x)=\cl{V_p(x)}\cap\cl{\wspc\setminus V_p(x)}$ denote the boundary of $V_p(x)$, where substituting~\eqref{eqn: Voronoi cell alternative} into the equality yields
\begin{equation}\label{eqn: Voronoi cell boundary}
    \partial V_p(x) = \bigcup_{q\in\verts\setminus\{p\}} \mathfrak{B}_{pq}(x)
\end{equation}
with $\mathfrak{B}_{pq}(x)\triangleq V_p(x)\cap\cl{\wspc\setminus V_{pq}(x)}$.
Note that the set $\mathfrak{B}_{pq}(x)$ denotes the interface between the Voronoi cells of agents $p$ and $q$. 
Moreover, $\mathfrak{B}_{pq}(x)$ can be shown to be orthogonal to the line segment between $x_p$ and $x_q$.

\begin{theorem} \label{thm: A_nu is attractive}
If $\tilde{\eta}_{\max}\in (0,k_1)$ and, for each $p\in\verts$, the timer parameters $T_1^p$, $T_2^p$ are selected to satisfy $0<T_1^p\leq T_2^p$ and the maximum dwell-time condition in~\eqref{eqn: Maximum dwell-time}, then every maximal solution $\phi$, with initial condition $\phi(0,0)\in\Phi$ and $\cl{\rge{\phi}}\subset U$, of the hybrid system $\hyb$ with data $(C,f,D,G)$ converges to the set $\attr$, i.e., the MAS achieves $\nu$-approximate coverage.
\hfill$\triangle$
\end{theorem}
\begin{proof}
Motivated by the locational cost in~\eqref{eqn: Locational Cost}, let $V\colon\statespc\to\RR_{\geq 0}$ be a Lyapunov-like function, such that 
\begin{equation} \label{eqn: V}
    V(\xi) \triangleq \sum_{p\in\verts} \int_{V_p(x)} \Vert x_p - z \Vert^2\varphi(z)\dd z = \sum_{p\in\verts} \locost_p(x).
\end{equation}
Observe that the Lyapunov-like function $V(\xi)$ is continuously differentiable.
Recalling that $\xi = (x,\eta,\tau)$, the gradient of $V(\xi)$ is given by
\begin{equation} \label{eqn: Partials of V}
\begin{aligned}
    \nabla V(\xi) &= \left[\frac{\partial V(s)}{\partial x}, \frac{\partial V(s)}{\partial \eta}, \frac{\partial V(s)}{\partial \tau} \right]_{s=\xi}^\top,\\
    \frac{\partial V(s)}{\partial x} &=\Bigg[ \sum_{p\in\verts}\frac{\partial\locost_p(s)}{\partial x_1}, \sum_{p\in\verts}\frac{\partial\locost_p(s)}{\partial x_2},..., \sum_{p\in\verts}\frac{\partial\locost_p(s)}{\partial x_N}\Bigg]\\
    \frac{\partial V(s)}{\partial \eta} &= 0_{nN}^\top, \quad \frac{\partial V(s)}{\partial \tau} = 0_N^\top.
\end{aligned}
\end{equation}
Since $\partial\locost_p/\partial x_q = 0_n^\top$ for all $p\notin\nbd_q(x)\cup\{q\}$, one has that
\begin{equation} \label{eqn: sum partial Lp by partial xq}
    \sum_{p\in\verts}\frac{\partial\locost_p}{\partial x_q} = \frac{\partial\locost_q}{\partial x_q} + \sum_{p\in\nbd_q(x)}\frac{\partial\locost_p}{\partial x_q}.  
\end{equation}
For $p\neq q\in\verts$, one can use~\eqref{eqn: Voronoi cell boundary} and the Leibniz integral rule to obtain
\begin{equation} \label{eqn: partial Lp by partial xq}
    \frac{\partial\locost_p}{\partial x_q} = \intop_{\mathfrak{B}_{pq}(x)}\Vert x_p-z \Vert^2 \varphi(z)\frac{(x_q - z)^\top}{\Vert x_q - x_p\Vert }{\tt d}z. 
\end{equation}
Similarly, for $p=q$, \eqref{eqn: Voronoi cell boundary} and the Leibniz integral rule enable the derivation of
\begin{equation} \label{eqn: partial Lq by partial xq}
\begin{aligned}
    \frac{\partial\locost_q}{\partial x_q} &= -\sum_{r\in\nbd_q(x)} \, \intop_{\mathfrak{B}_{qr}(x)}\Vert x_q-z \Vert^2 \varphi(z)\frac{(x_q - z)^\top}{\Vert x_r - x_q\Vert }{\tt d}z\\
    &\phantom{=} + \intop_{V_q(x)}2(x_q - z)^\top \varphi(z){\tt d}z.
\end{aligned}
\end{equation}
For each pair of agents $p,q\in\verts$ with $p\in\nbd_q$, one can show that $\mathfrak{B}_{pq}(x) = \mathfrak{B}_{qp}(x)$ and $\Vert x_p - z \Vert = \Vert x_q - z\Vert$ for $z\in\mathfrak{B}_{pq}$, where the second equality follows from the fact that $\mathfrak{B}_{pq}(x)$ bisects to the line segment defined by $x_p$ and $x_q$.
The substitution of~\eqref{eqn: partial Lp by partial xq} and~\eqref{eqn: partial Lq by partial xq} into~\eqref{eqn: sum partial Lp by partial xq} and the replacement of the summation index $r$ with $p$ yields
\begin{equation}
    \sum_{p\in\verts}\frac{\partial\locost_p}{\partial x_q} =  \intop_{V_q(x)}2(x_q - z)^\top \varphi(z){\tt d}z = -2m_q(x) e_q^\top,
\end{equation}
where the second equality follows from the substitution of~\eqref{eqn: Centroid}--\eqref{eqn: Tracking error}.
Therefore, the partial derivative of $V(\xi)$ with respect to $x$ is
\begin{equation} \label{eqn: partial V wrt partial x}
    \left.\frac{\partial V(s)}{\partial x}\right\vert_{s=\xi} = -2\left[m_1(x) e_1^\top, m_2(x) e_2^\top,...,m_N(x) e_N^\top\right].
\end{equation}

When $\xi\in C$, the change in $V(\xi)$ is computed using $\dot{V}(\xi)=\langle\nabla V(\xi),f(\xi)\rangle$, where $f(\xi)$ denotes the single-valued flow map in~\eqref{eqn: Flow map}.
Substituting~\eqref{eqn: Flow map}, \eqref{eqn: Partials of V}, and~\eqref{eqn: partial V wrt partial x} into $\langle\nabla V(\xi),f(\xi)\rangle$ yields
\begin{equation} \label{eqn: VDot One}
    \dot{V}(\xi) = \frac{\partial V}{\partial x}\dot{x} + \frac{\partial V}{\partial\eta}\dot{\eta} + \frac{\partial V}{\partial\tau}\dot{\tau} = -2\sum_{p\in\verts} m_p(x)e_p^\top\eta_p.
\end{equation}
The substitution of~\eqref{eqn: eta_p tilde} into~\eqref{eqn: VDot One} yields
\begin{equation} \label{eqn: VDot Two}
    \dot{V}(\xi) = 2\sum_{p\in\verts}m_p(x)\left(-e_p^\top\tilde{\eta}_p - k_1 e_p^\top\sat{e_p,\tilde{\nu}}\right).
\end{equation}
Recall that $\Vert \tilde{\eta}_p \Vert\leq \tilde{\eta}_{\max}$ for every $p\in\verts$ since each parameter $T_2^p$ satisfies the maximum dwell-time condition in~\eqref{eqn: Maximum dwell-time}.
Hence, the time derivative of $V(\xi)$ in~\eqref{eqn: VDot Two} can be bounded as
\begin{equation} \label{eqn: VDot Three}
    \dot{V}(\xi) 
    \leq 2\sum_{p\in\verts}m_p(x)\left( \tilde{\eta}_{\max}\Vert e_p\Vert  - k_1 e_p^\top\sat{e_p,\tilde{\nu}}\right).
\end{equation}
Recall $\tilde{\eta}_{\max}\in (0, k_1)$, which yields $\tilde{\eta}_{\max}\tilde{\nu}/k_1 < \tilde{\nu} < \nu$.
Since the output of $\sat{e_p,\tilde{\nu}}$ depends on the value of $\Vert e_p \Vert$, we now analyze each summand of~\eqref{eqn: VDot Three} individually, where we restrict our attention to when $\xi\in C\cap U$.
\textbf{Case I}: $\Vert e_p \Vert > \tilde{\nu}$.
Then,
\begin{equation} \label{eqn: Summand Case I}
    \tilde{\eta}_{\max}\Vert e_p\Vert  - k_1 e_p^\top\sat{e_p,\tilde{\nu}} = -(k_1 - \tilde{\eta}_{\max})\Vert e_p \Vert < 0.  
\end{equation}
\textbf{Case II}: $\Vert e_p \Vert \in [\tilde{\eta}_{\max}\tilde{\nu}/k_1,\tilde{\nu}]$.
Then,
\begin{equation} \label{eqn: Summand Case II}
    \tilde{\eta}_{\max}\Vert e_p\Vert  - k_1 e_p^\top\sat{e_p,\tilde{\nu}} = -\frac{k_1}{\tilde{\nu}}\left(\Vert e_p \Vert - \frac{\tilde{\eta}_{\max}\tilde{\nu}}{k_1} \right) \Vert e_p \Vert \leq 0.  
\end{equation}
Therefore, when $\xi\in C\cap U$, the inequalities in~\eqref{eqn: VDot Three}--\eqref{eqn: Summand Case II} yield
\begin{equation} \label{eqn: VDot Four}
    \dot{V}(\xi) 
    \leq 2\sum_{p\in\verts}m_p(x)\left( \tilde{\eta}_{\max}\Vert e_p\Vert  - k_1 e_p^\top\sat{e_p,\tilde{\nu}}\right) \leq 0.
\end{equation}

When $\xi\in D$ and $g\in G(\xi)$, the change in $V(\xi)$ is computed using $V(g) - V(\xi)$.
Since $V(\xi)$ in~\eqref{eqn: V} is a continuous function of $\xi$, $V(g) - V(\xi)= 0$ whenever $\xi\in D$.
Recall the definitions of $u_C(z)$ and $u_D(z)$ in Section~\ref{sec: hybrid systems}.
Using~\eqref{eqn: VDot Four}, one has that $u_C(\xi) \leq 0$ for all $\xi\in C\cap U$.
Similarly, $u_D(\xi) = 0$ for all $\xi\in D\cap U$ since $V(g) - V(\xi) = 0$ whenever $\xi\in D$ and $g\in G(\xi)$.
Therefore, the invariance principle in~\cite[Corollary 8.4]{Goebel.Sanfelice.ea2012} implies that, for some $r\in V(U)$, every precompact solution $\phi^\ast$ of $\hyb$ such that $\cl{\rge{\phi^\ast}}\subset U$ approaches the largest weakly invariant subset of
\begin{equation} \label{eqn: largest weakly invariant set}
    V^{-1}(r)\cap U\cap \Big[\cl{u_C^{-1}(0)}\cup \big(u_D^{-1}(0)\cap G(u_D^{-1}(0))\big)\Big].
\end{equation}
Fix a maximal solution $\phi$ of $\hyb$ with initial condition $\phi(0,0)\in\Phi$ and $\cl{\rge{\phi}}\subset U$.
We now show this $\phi$ is precompact.
By Lemma~\ref{lemma: Completeness}, all maximal solutions of $\hyb$ are complete.
Along the maximal solution $\phi$, the trajectories $\eta(\phi(t,j))$ and $\tau(\phi(t,j))$ are bounded by construction for all $(t,j)\in\dom{\phi}$.
Since $\phi$ is maximal, $\cent_p(x(\phi(t,j)))$ is well-defined for every $p\in\verts$ and all $(t,j)\in\dom{\phi}$.
For every $p\in\verts$, $\dot{x}_p=\eta_p$, $\eta_p$ is a sample-and-hold approximation of $k_1\sat{e_p,\tilde{\nu}}$, $e_p = \cent_p(x) - x_p$, the length of the flow intervals of agent $p$ are bounded by $T_2^p$, and the centroid $\cent_p(x)$ is contained in $\wspc$.
Hence, there exists a bounded \fred{set} $M$ \fred{containing} $\wspc^N$ such that $x(\phi(t,j))\in M$ for all $(t,j)\in\dom{\phi}$,
implying that $x(\phi(t,j))$ is bounded for all $(t,j)\in\dom{\phi}$.
Consequently, $\phi$ is precompact.
Using~\eqref{eqn: Summand Case II}, \eqref{eqn: VDot Four}, and the \fred{definitions of $u_C(z)$ and} $\attr$, one can see that 
\begin{equation*}
\begin{aligned}
    \phantom{=}& \ U\cap \cl{u_C^{-1}(0)} = U\cap\{\xi\in C\colon u_C(\xi) = 0 \}\\
    =& \ \{\xi\in C\cap U \colon \forall_{p\in\verts} \ \Vert e_p \Vert = \tilde{\eta}_{\max}\tilde{\nu}/k_1 \}\\
    \subset& \ \{\xi\in C\cap U \colon \forall_{p\in\verts} \ \Vert e_p \Vert \leq \nu \} \subset \attr.
\end{aligned}
\end{equation*}
In addition, $u_D^{-1}(0) = D$, implying that $u_D^{-1}(0)\cap G(u_D^{-1}(0))=\varnothing$.
Hence, the set in~\eqref{eqn: largest weakly invariant set} is contained within $V^{-1}(r)\cap\attr$, and solutions of $\hyb$ converge to $\attr$.
\end{proof}

\section{Simulation Examples}
This section presents the results from four numerical simulations.
In the first simulation, we validate our coverage controller on a $30$-agent MAS.
Next, using a $12$-agent MAS, we simulate the continuous-time Lloyd algorithm presented in~\cite{Cortes.Martinez.ea2004}, the self-triggered coverage controller from~\cite{RodriguezSeda.Xu.ea2023}, and our result.
We end the section with a discussion of the results from the last three simulations.
Note, the same workspace and density function are used in all simulations.
All four simulations were conducted in MATLAB.

\subsection{Validation of the Timer-Based Coverage Controller}
\begin{figure}[t]
\centering
    \begin{subfigure}[b]{0.49\columnwidth}
         \centering
         \includegraphics[width=1\columnwidth]{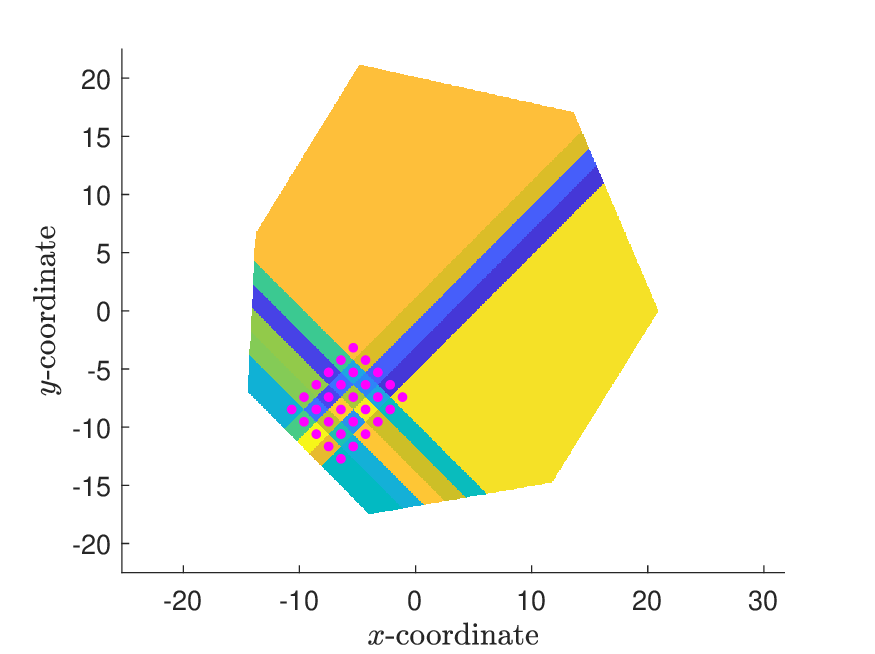}
    \end{subfigure}
    \begin{subfigure}[b]{0.49\columnwidth}
         \centering
         \includegraphics[width=1\columnwidth]{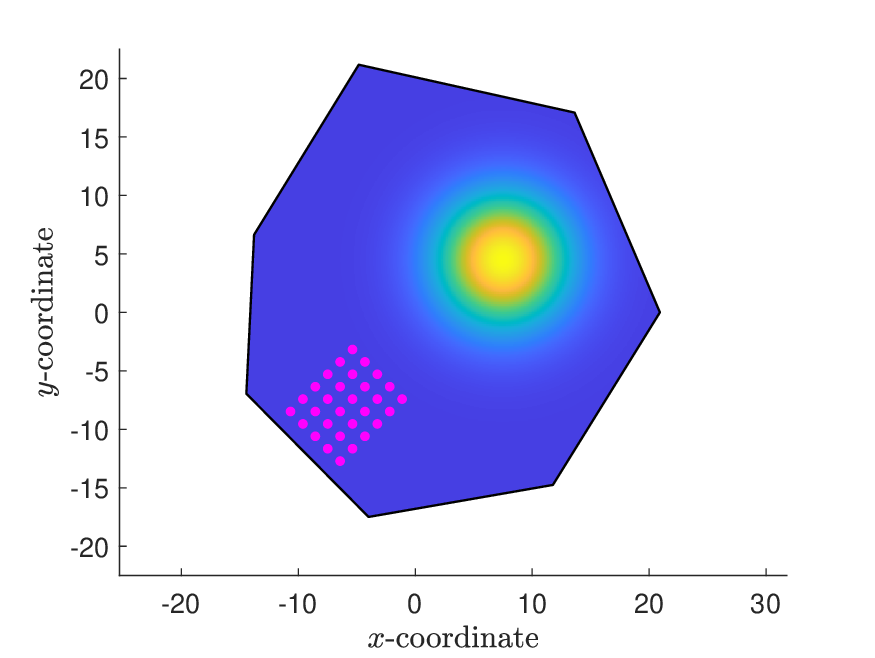} 
    \end{subfigure}
    \begin{subfigure}[b]{0.49\columnwidth}
         \centering
         \includegraphics[width=1\columnwidth]{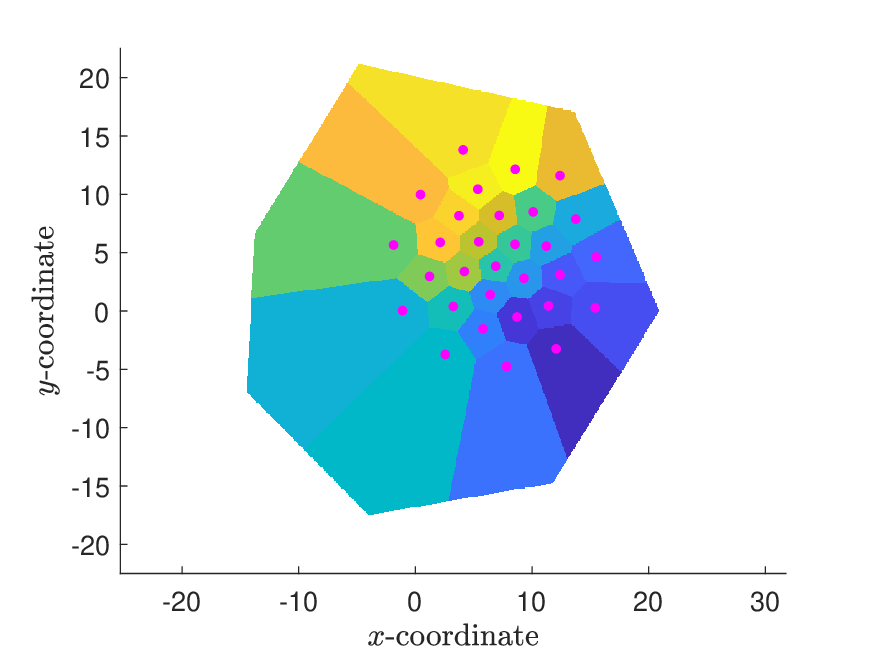}
    \end{subfigure}
    \begin{subfigure}[b]{0.49\columnwidth}
         \centering
         \includegraphics[width=1\columnwidth]{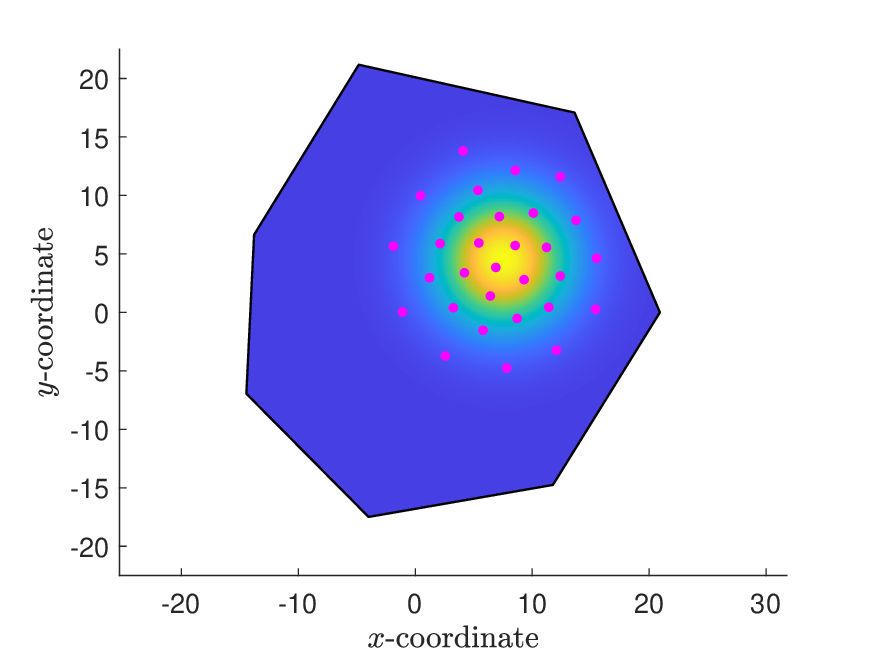} 
    \end{subfigure}
    \caption{
    \textbf{(Top Left)} Illustration of the heptagonal workspace $\wspc$ overlaid with the initial MAS configuration (magenta dots) and the corresponding Voronoi tessellation of $\wspc$ using the initial MAS configuration as a generator.
    \textbf{(Bottom Left)} Illustration of the heptagonal workspace $\wspc$ overlaid with the final MAS configuration (magenta dots) and the corresponding Voronoi tessellation of $\wspc$ using the final MAS configuration as a generator.
    \textbf{(Top Right)} Depiction of the workspace $\wspc$, the heat map resulting from the density function $\varphi(s)$, and the initial MAS configuration.
    The most important points in the workspace are those closest to the peak of the Gaussian density, colored in yellow.
    \textbf{(Bottom Right)} Depiction of the workspace $\wspc$, the heat map resulting from the density function $\varphi(s)$, and the final MAS configuration.}
    \label{fig: configurations}
\end{figure}

\begin{figure}
\centering
\includegraphics[width=1\columnwidth]{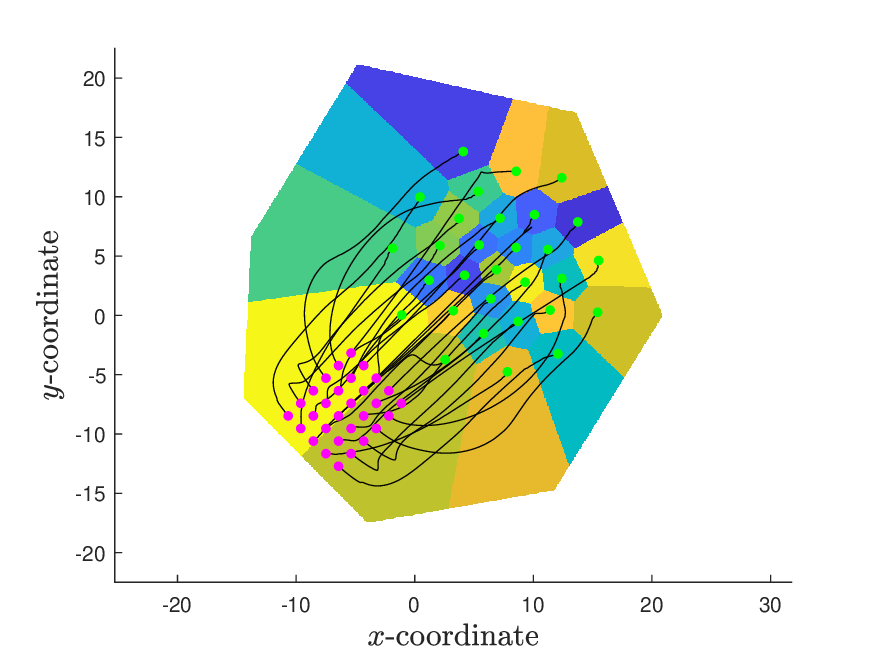} 
\caption{
Image of the workspace $\wspc$, the initial MAS configuration in magenta dots, the final MAS configuration in green dots, the Voronoi tessellation of $\wspc$ using the final MAS configuration as a generator, and the trajectories of all agent positions in black curves.
}
\label{fig: MAS trajectories}
\end{figure}

\begin{figure}
\centering
    \includegraphics[width=1\columnwidth]{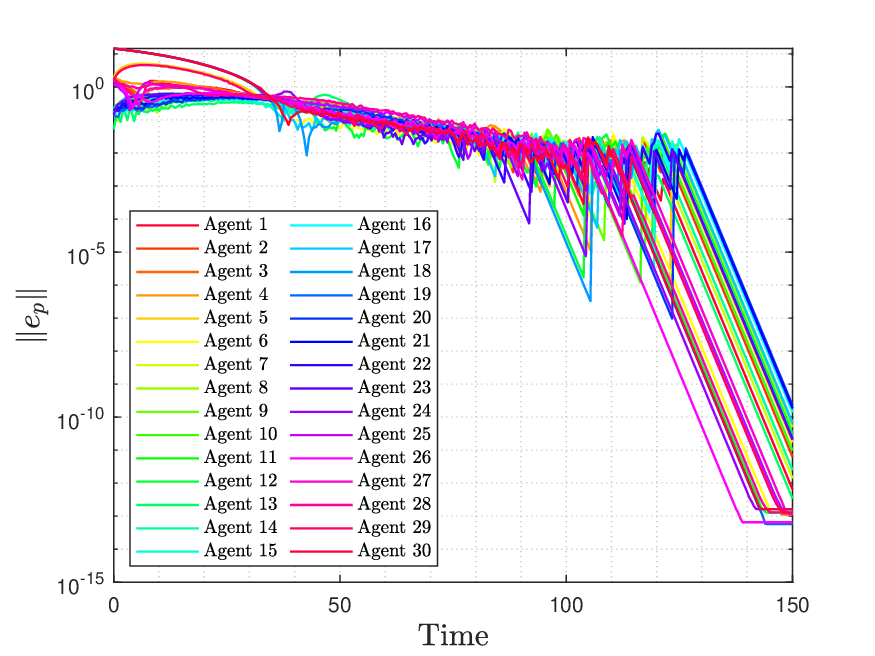} 
    \caption{
    Depiction of the point-wise Euclidean norm of the centroid tracking error $e_p$, defined in~\eqref{eqn: Tracking error}, for each agent $p\in\verts$ versus time.
    The horizontal and vertical axes are in linear and logarithmic scales, respectively.
    Since the magnitudes of the centroid tracking errors are bounded by $0.7$ for all time $t\geq 60$, the MAS achieves $\nu$-approximate coverage with the desired $\nu$ of $0.7$. 
    Moreover, the timer-based coverage controller provided better error regulation than afforded by the theory, speaking to the conservatism of the result.}
    \label{fig: tracking errors}
\end{figure}

\begin{figure}
\centering
    \includegraphics[width=1\columnwidth]{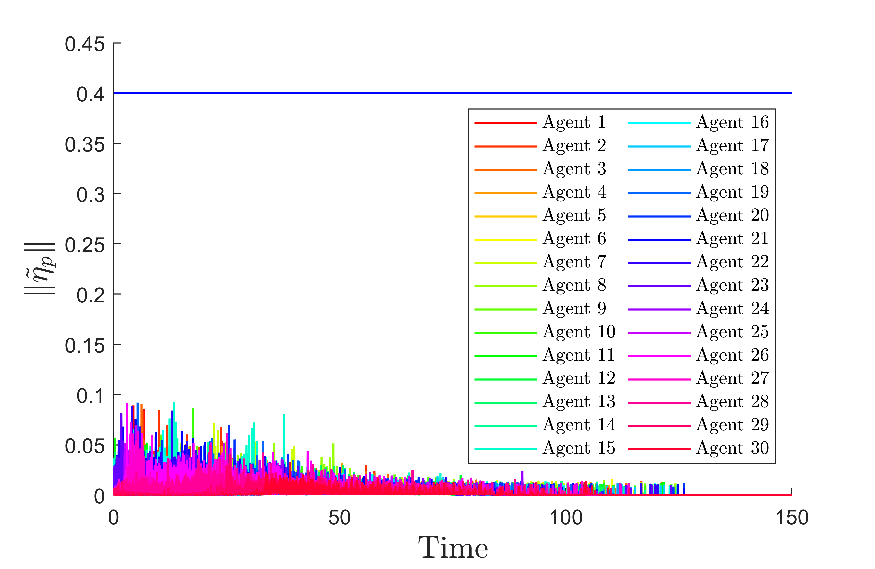} 
    \caption{
    Portrayal of the point-wise Euclidean norm of the error $\tilde{\eta}_p$, defined in~\eqref{eqn: eta_p tilde}, for each agent $p\in\verts$ versus time.
    The horizontal blue line represents the equation $y=\tilde{\eta}_{\max}$.
    Under the selected simulation parameters, one can see that $\Vert \fred{\tilde{\eta}_p} \Vert \leq \tilde{\eta}_{\max}$ for all time $t\in[0,150]$, as guaranteed by the maximum dwell-time condition.}
    \label{fig: EtaTilde}
\end{figure}

\begin{figure}
\centering
    \includegraphics[width=1\columnwidth]{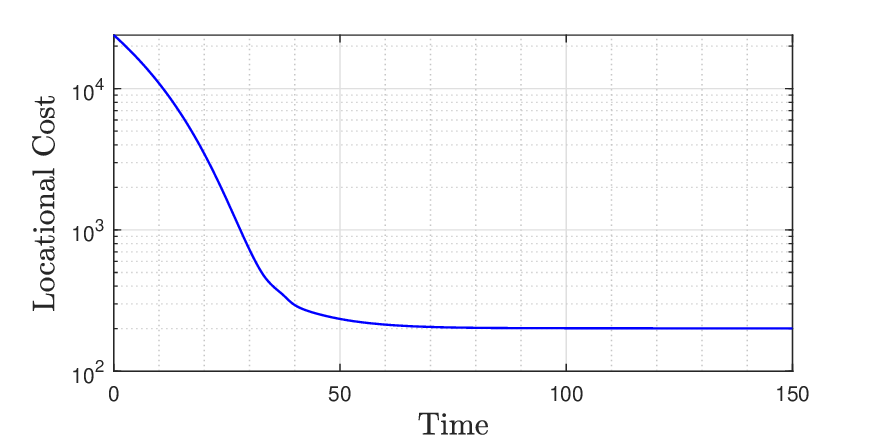} 
    \caption{
    Illustration of the locational cost in~\eqref{eqn: Locational Cost} versus time, where the horizontal and vertical axes are in linear and logarithmic scales, respectively.
    The steady state value of the locational cost is about $201$.}
    \label{fig: locational cost}
\end{figure}

\begin{figure}
\centering
    \includegraphics[width=1\columnwidth]{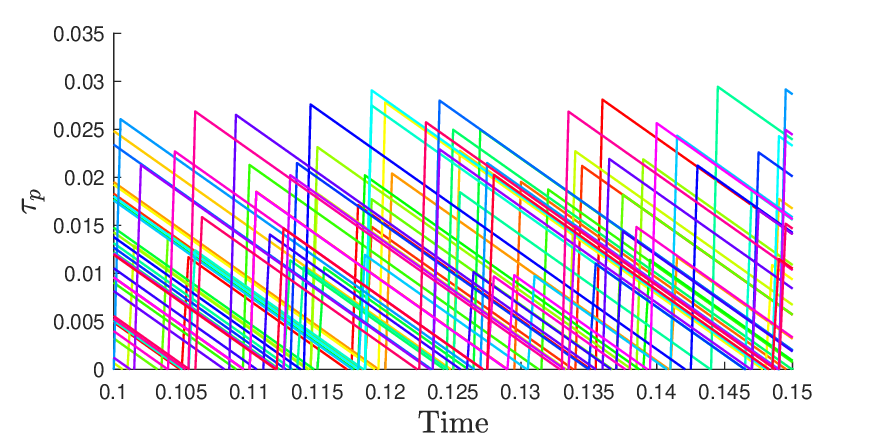} 
    \caption{
    Depiction of the evolution of the timer variables in $\{\tau_p\}_{p\in\verts}$ versus time for $t\in[0.1,0.15]$.
    When $\tau_p$ equals zero, an event occurs for agent $p$, which includes the resetting of $\tau_p$ to a value in $[T_1^p,T_2^p]$.
    Since the timers reach zero at different times and the distance between consecutive event times for each agent is positive and varying, events between agents occur intermittently and asynchronously.
    Note that the curve colors match the legend description in Figures~\ref{fig: tracking errors} and ~\ref{fig: EtaTilde}.}
    \label{fig: timer plot}
\end{figure}

The coverage controller defined by $u_p=\eta_p$, \eqref{eqn: Timer}, and~\eqref{eqn: eta_p} for each agent $p\in \verts$ was simulated with the following parameters: $n=2$, $N=30$, $\tilde{\eta}_{\max}=0.4$, $k_1=0.525$, $\varepsilon=10^{-8}$, $\nu=0.7$, $L_p=5$, $T_1^p=0.01$, and $T_2^p = 0.03$.
For every $p,q\in\verts$, we have $L_p=L_q$, $T_1^p=T_1^q$, and $T_2^p=T_2^q$ .
A Gaussian density function, specified by $\varphi(s) \triangleq \exp(-0.03\Vert s-c\Vert^2)$ centered about $c=[7.5, 4.5]^\top$, was defined on a heptagonal workspace $\wspc$.
Figures~\ref{fig: configurations}--\ref{fig: timer plot} depict the coordinates of the workspace vertices, initial configuration of the MAS, and simulation results.
With respect to the initial condition of $\eta$ and $\tau$, we used $\eta_p(0,0)=k_1\sat{e_p(0,0),\tilde{\nu}}$ and $\tau_p(0,0)=T_2^p$ for each $p\in\verts$ so that $\phi(0,0)\in\Phi$ was satisfied.
According to Theorem~\ref{thm: A_nu is attractive}, all centroid tracking errors should converge to a closed ball centered about the origin with radius $\nu = 0.7$.
Figure~\ref{fig: tracking errors} indicates the achievement of $\nu$-approximate coverage with $\nu= 10^{-9}$, which is better than expected.

\subsection{Continuous-Time Lloyd Algorithm}
\begin{figure}
\centering
    \begin{subfigure}[b]{0.49\columnwidth}
         \centering
         \includegraphics[width=0.95\columnwidth]{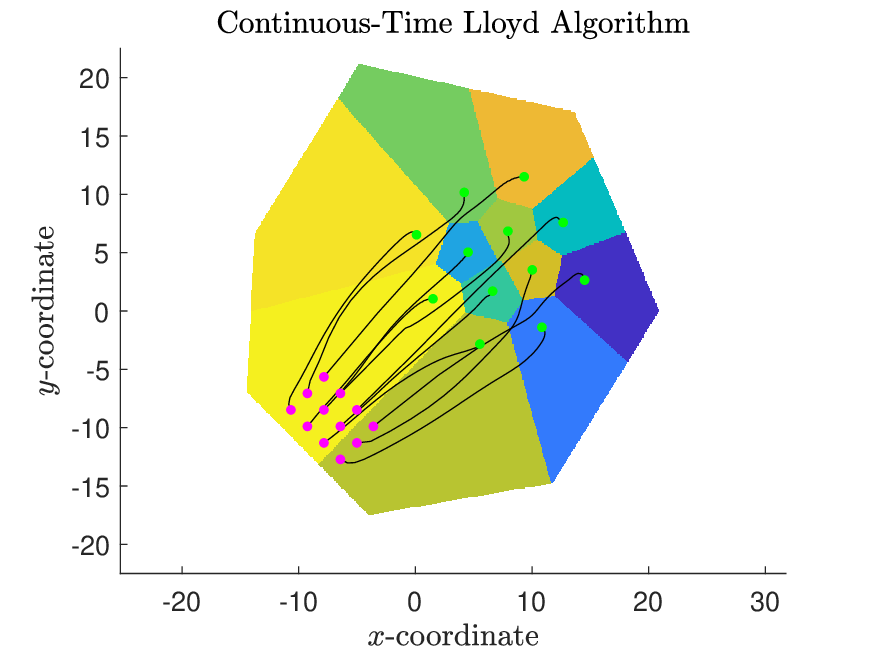}
    \end{subfigure}
    \begin{subfigure}[b]{0.49\columnwidth}
         \centering
         \includegraphics[width=0.95\columnwidth]{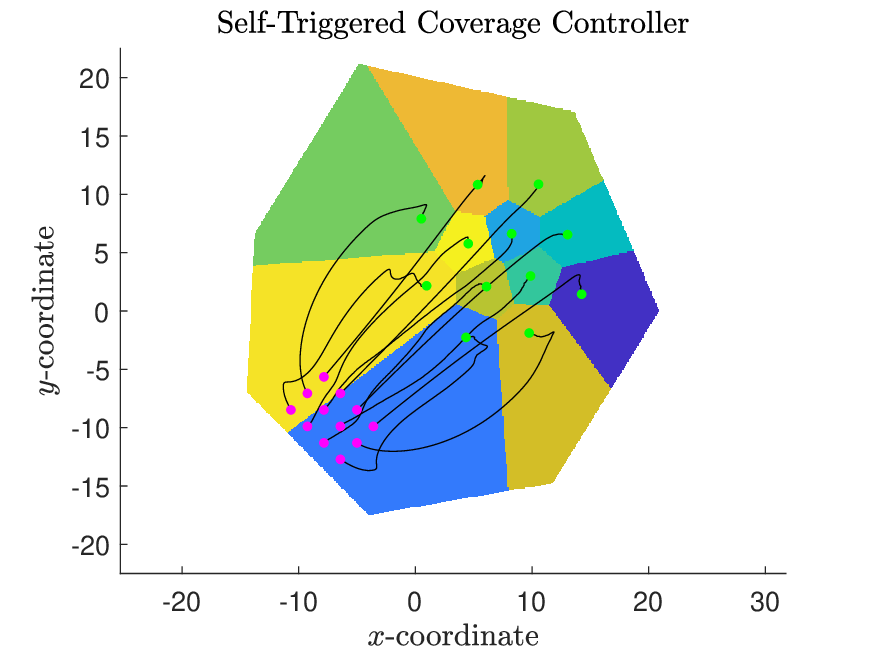}
    \end{subfigure}
    \begin{subfigure}[b]{0.49\columnwidth}
         \centering
         \includegraphics[width=0.95\columnwidth]{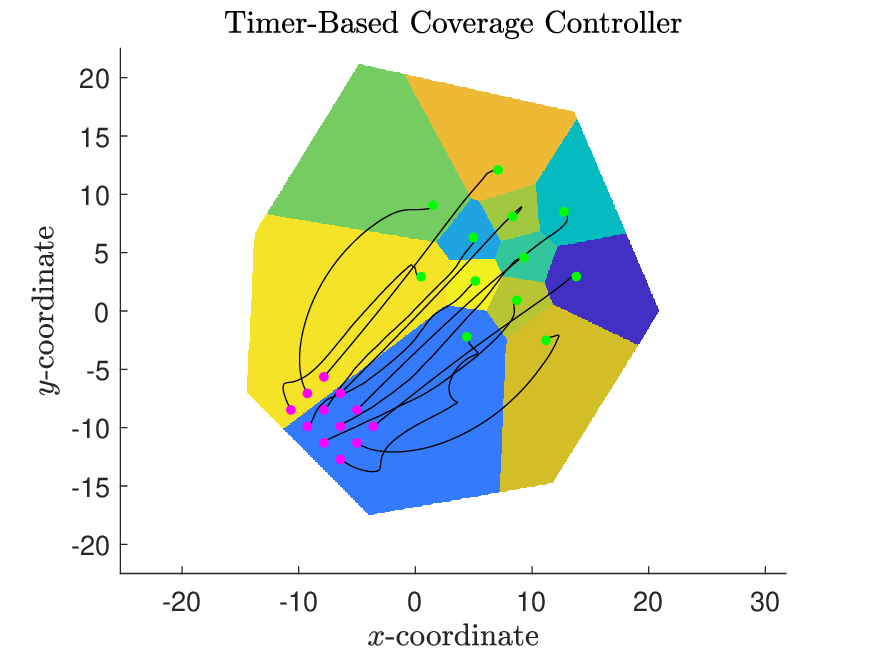}
    \end{subfigure} 
    \caption{
    Each image illustrates the workspace $\wspc$, the initial MAS configuration in magenta dots, the final MAS configuration in green dots, the Voronoi tessellation of $\wspc$ using the final MAS configuration as a generator, and the trajectories of all agent positions in black curves for the simulations of the continuous-time Lloyd algorithm, the self-triggered coverage controller, and the timer-based coverage controller.}
    \label{fig: MAS trajectories Comparison}
\end{figure}

\begin{figure}
\centering
    \begin{subfigure}[b]{1\columnwidth}
         \centering
         \includegraphics[width=1\columnwidth]{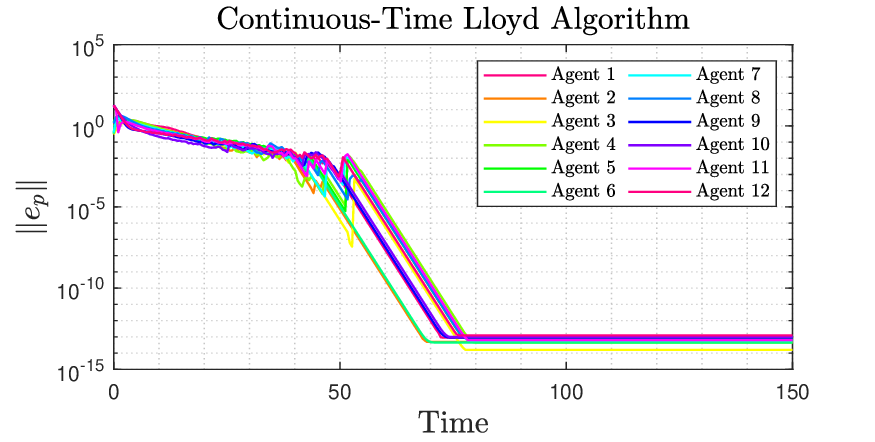}
    \end{subfigure}
    \begin{subfigure}[b]{1\columnwidth}
         \centering
         \includegraphics[width=1\columnwidth]{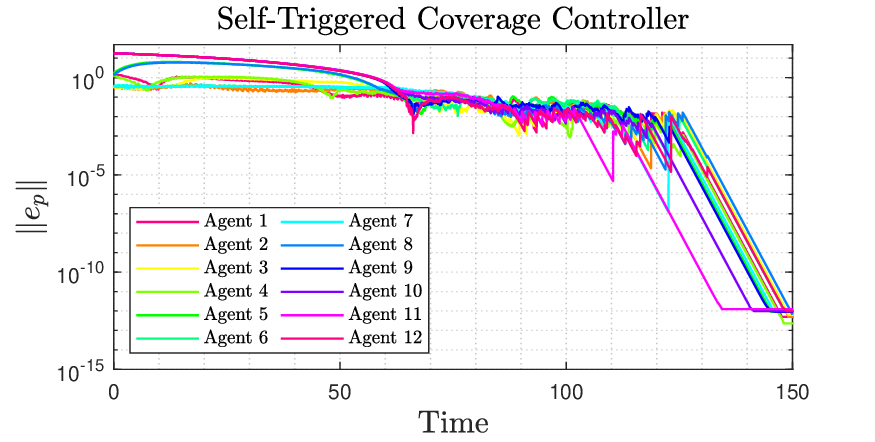}
    \end{subfigure}
    \begin{subfigure}[b]{1\columnwidth}
         \centering
         \includegraphics[width=1\columnwidth]{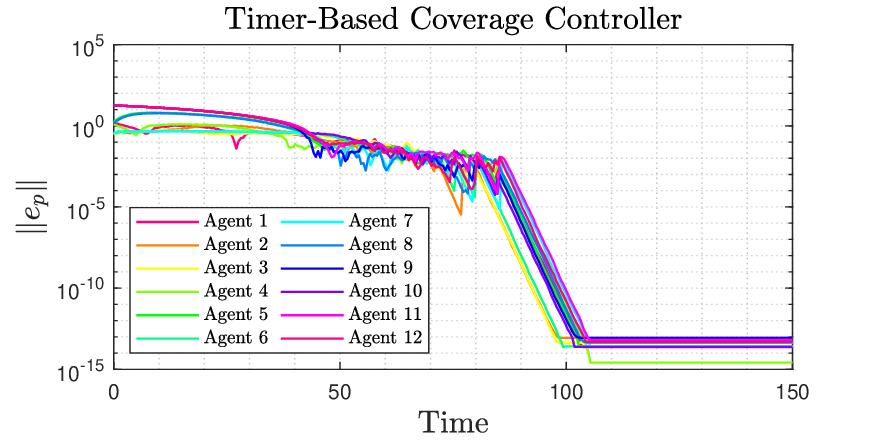}
    \end{subfigure} 
    \caption{
    Each image shows the point-wise Euclidean norm of the centroid tracking error $e_p$ for each $p\in\verts$ versus time.
    The horizontal and vertical axes are in linear and logarithmic scales, respectively.
    The simulation results are for the continuous-time Lloyd algorithm, the self-triggered coverage controller, and the timer-based coverage controller.}
    \label{fig: tracking errors Comparison}
\end{figure}

\begin{figure}
\centering
    \begin{subfigure}[b]{0.49\columnwidth}
         \centering
         \includegraphics[width=0.9\columnwidth]{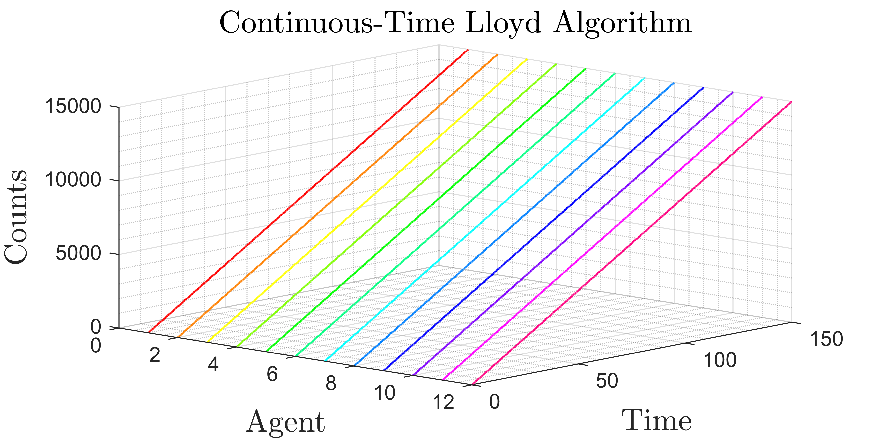}
    \end{subfigure}
    \begin{subfigure}[b]{0.49\columnwidth}
         \centering
         \includegraphics[width=0.9\columnwidth]{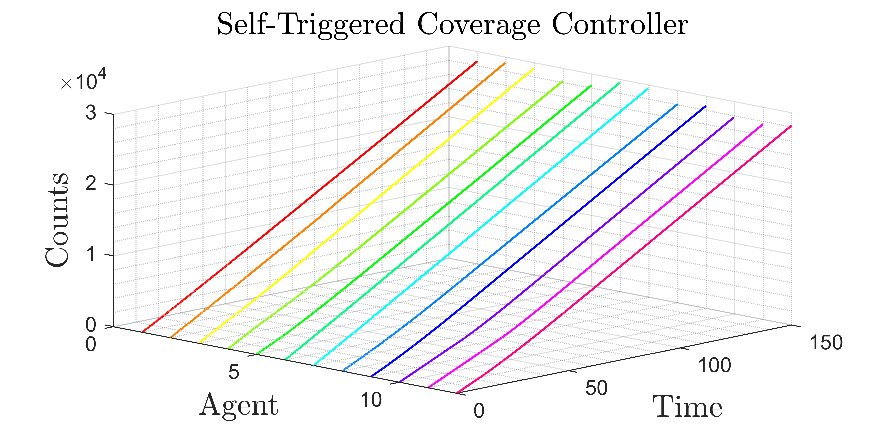}
    \end{subfigure}
    \begin{subfigure}[b]{0.49\columnwidth}
         \centering
         \includegraphics[width=0.9\columnwidth]{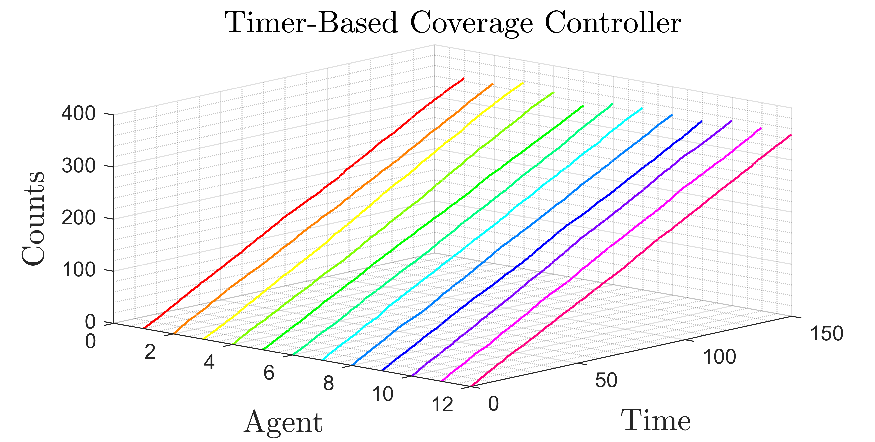}
    \end{subfigure} 
    \caption{
    Each plot shows the cumulative number of times agent $p$, for each $p\in\verts$, computed Voronoi-cell-like objects as a function of time.
    The simulation results are for the continuous-time Lloyd algorithm, the self-triggered coverage controller, and the timer-based coverage controller.}
    \label{fig: Voronoi Count Comparison}
\end{figure}

The controller of agent $p$ under the continuous-time Lloyd algorithm in~\cite[Equation 6]{Cortes.Martinez.ea2004} can be written as $u_p = k_2 e_p$, where $k_2\in\RR_{>0}$ is a user-defined parameter and $e_p$ is the centroid tracking error in~\eqref{eqn: Tracking error}.
The simulation parameters are $n=2$, $N=12$, and $k_2 = 1$.
The subplots labeled "Continuous-Time Lloyd algorithm" in Figures~\ref{fig: MAS trajectories Comparison}--\ref{fig: Voronoi Count Comparison} show the simulation results.

\subsection{Self-Triggered Coverage Control}
\begin{figure}[t]
\centering
    \begin{subfigure}[b]{0.49\columnwidth}
         \centering
         \includegraphics[width=0.95\columnwidth]{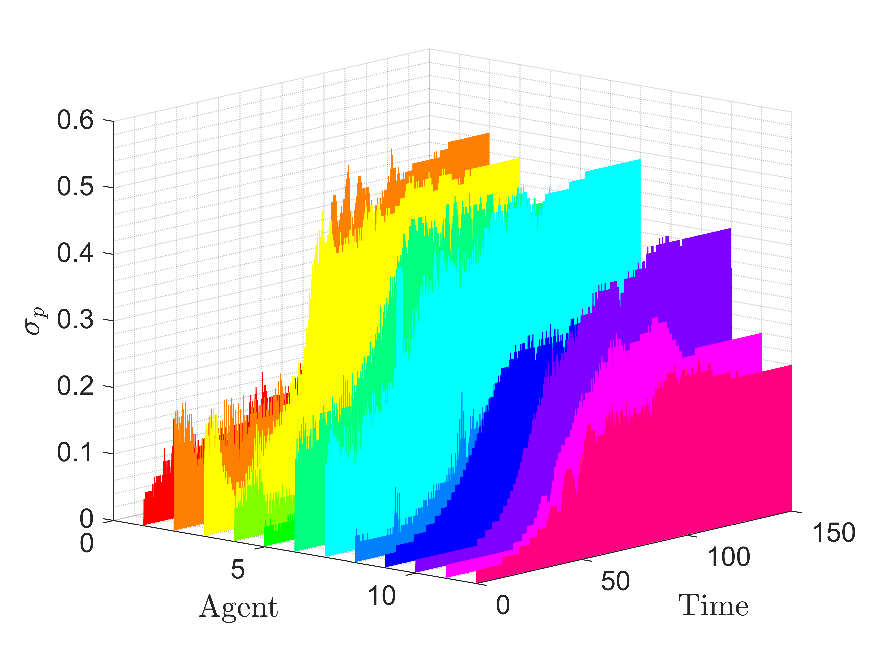}
    \end{subfigure}
    \begin{subfigure}[b]{0.49\columnwidth}
         \centering
         \includegraphics[width=0.95\columnwidth]{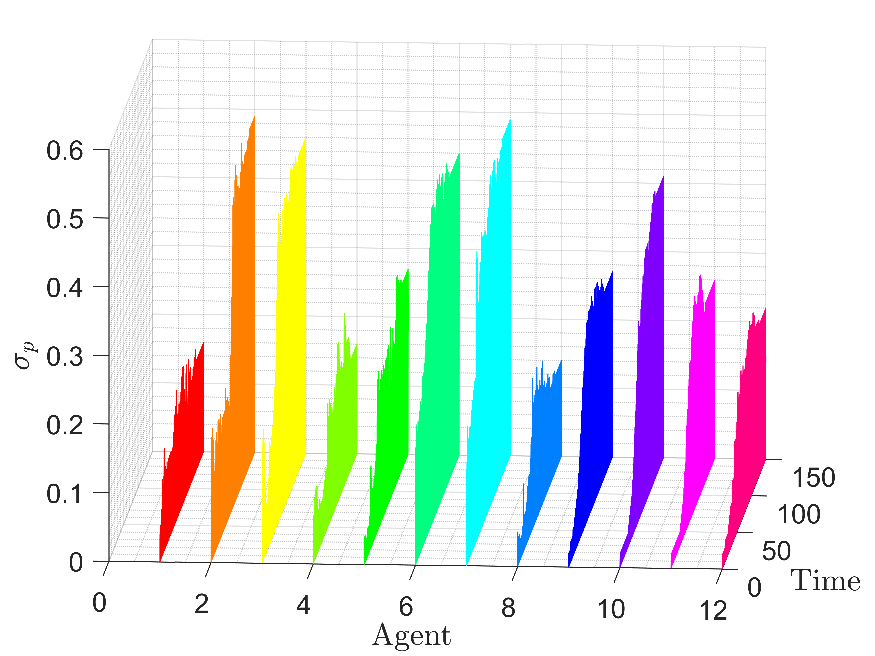} 
    \end{subfigure}
    \begin{subfigure}[b]{0.49\columnwidth}
         \centering
         \includegraphics[width=0.95\columnwidth]{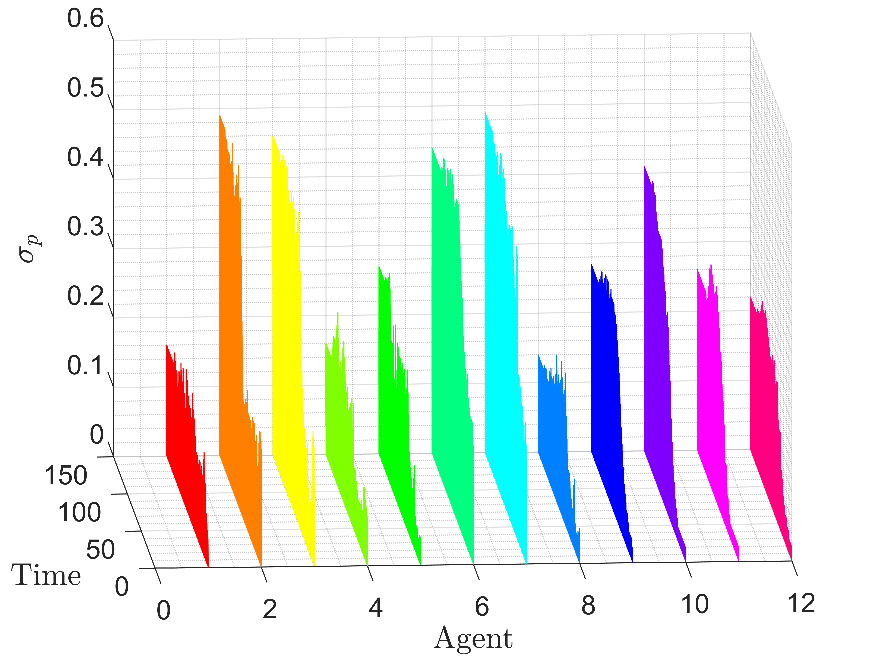}
    \end{subfigure}
    \begin{subfigure}[b]{0.49\columnwidth}
         \centering
         \includegraphics[width=0.95\columnwidth]{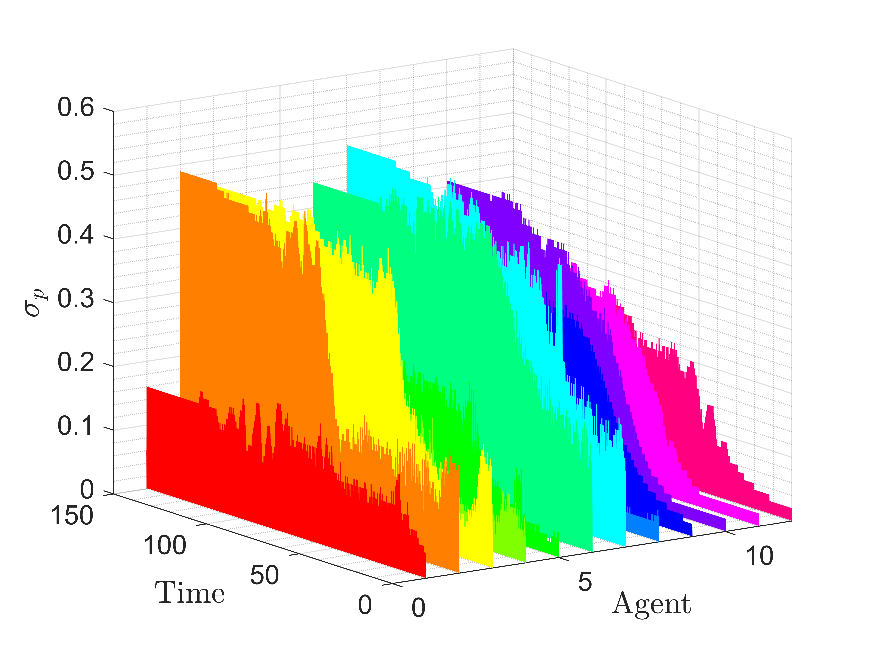} 
    \end{subfigure}
\caption{
Illustrations of the evolution of the timer $\sigma_p$ in~\eqref{eqn: self trigger timer} versus time for each agent $p\in\verts$.
As the simulation progresses, the inter-event times of some agents tend to increase, which reduces the amount of communication/sensing needed to acquire position information.
The results are for the self-triggered coverage controller.}
\label{fig: timer plots Self Trigger}
\end{figure}

Using~\cite[Equation 5]{RodriguezSeda.Xu.ea2023}, the self-triggered coverage controller of agent $p$ can be expressed as $u_p = \kappa\cdot\text{sat}_{v/\kappa}(\cent_p(x(t_k^p)) - x_p)$, where $\kappa, v\in\RR_{>0}$ are user-defined parameters, $\text{sat}_{\mathtt{a}}(s) = s$ if $\Vert s \Vert \leq \mathtt{a}$ and $\text{sat}_{\mathtt{a}}(s) = \mathtt{a} \cdot s/\Vert s \Vert$ if $\Vert s \Vert > \mathtt{a}$ for a vector $s$ and positive constant $\mathtt{a}$, and $\{t_k^p\}_{k=0}^\infty$ denotes the sequence of event times of agent $p$ generated by the self trigger mechanism in~\cite[Equation 8]{RodriguezSeda.Xu.ea2023}.
Unfortunately, \cite[Assumption 1]{RodriguezSeda.Xu.ea2023} cannot be verified \textit{a priori}.
As a result, we cannot satisfy the sufficient conditions of~\cite[Corollary 1]{RodriguezSeda.Xu.ea2023}, which provide guarantees for 1) the maximal distance between the configuration of the MAS and a CVC at steady state given~\cite[Proposition 4]{RodriguezSeda.Xu.ea2023} and 2) the minimum inter-event time of each agent being lower bounded by the user-defined parameter $\tau_{\min}$.
Nevertheless, we selected parameters that enabled comparable performance relative to the continuous-time Lloyd algorithm.
The simulation parameters are $n=2$, $N=12$, $\kappa = 1$, $v=0.35$, $\tau_{\min} = 0.01$, $\tau_{\max} = 0.75$, and $\delta = 3$.
The subplots labeled "Self-Triggered Coverage Controller" in Figures~\ref{fig: MAS trajectories Comparison}--\ref{fig: Voronoi Count Comparison} and Figure~\ref{fig: timer plots Self Trigger} show the simulation results, which were generated by implementing~\cite[Algorithm 1]{RodriguezSeda.Xu.ea2023}.
Since the self trigger of an agent can create dynamism in the inter-event times of said agent, we utilize a timer mechanism to help visualize this phenomenon.
Specifically, for each $p\in\verts$, let $\sigma_p\in[0,\tau_{\max}]$ be a timer for agent $p$, such that 
\begin{equation} \label{eqn: self trigger timer}
\begin{aligned}
    \dot{\sigma}_p &= 1, & h_p(\sigma_p) &\leq 0 \text{ and } \sigma_p\in [0,\tau_{\max}] \\
    \sigma_p^+ &= 0, & h_p(\sigma_p) &\geq 0 \text{ or } \sigma_p\geq \tau_{\max}.
\end{aligned}
\end{equation}
Note, $\sigma_p(0) = 0$ and $h_p$ is the self-trigger function of agent $p$ defined in~\cite[Equation 8]{RodriguezSeda.Xu.ea2023}.
The value of $\sigma_p$ before each reset represents the length of the preceding flow interval.

\subsection{Timer-Based Coverage Control}
The coverage controller defined by $u_p=\eta_p$, \eqref{eqn: Timer}, and~\eqref{eqn: eta_p} for each agent $p\in \verts$ was simulated with the following parameters: $n=2$, $N=12$, $\tilde{\eta}_{\max}= 0.3$, $k_1=0.5$, $\varepsilon=10^{-8}$, $\nu=0.5$, $L_p=5$, $T_1^p = 0.2$, and $T_2^p = 0.65$.
For all $p,q\in\verts$, we utilize $L_p=L_q$, $T_1^p=T_1^q$, and $T_2^p=T_2^q$ .
For this simulation only, the selection of $T_2^p$ was not restricted by the maximum dwell-time condition for any agent $p\in\verts$.
This selection was made to establish the conservative nature of the maximum dwell-time condition.
The subplots labeled "Timer-Based Coverage Controller" in Figures~\ref{fig: MAS trajectories Comparison}--\ref{fig: Voronoi Count Comparison} show the simulation results.

\subsection{Discussion}
The results from the continuous-time Lloyd algorithm serve as a benchmark. 
As shown by Figures~\ref{fig: MAS trajectories Comparison} and~\ref{fig: tracking errors Comparison}, the continuous-time Lloyd algorithm was able to drive the $12$-agent MAS to a CVC, where the centroid tracking error of each agent may be bounded above by $1.23\times 10^{-13}$ for all $t\geq 80$. 
Since the maximum simulation time was $150$ and the minimum timestep was $0.01$, the continuous-time Lloyd algorithm had each agent compute their Voronoi cell a total of $15,000$ times as shown in Figure~\ref{fig: Voronoi Count Comparison}.
Note, each time agent $p$ computes its Voronoi cell which is needed to update its controller, agent $p$ must first measure its position and obtain the positions of its Voronoi neighbors via communication or sensing.
Hence, if $\zeta_p\in\RR_{>0}$ quantifies the fixed cost of agent $p$ computing its Voronoi cell and $n_p\in\ZZ_{>0}$ denotes the number of times agent $p$ computes its Voronoi cell, then the total computational cost of implementing the continuous-time Lloyd algorithm for agent $p$ is approximately $\zeta_p\cdot n_p$, i.e., $15,000\cdot \zeta_p$.

Given Figures~\ref{fig: MAS trajectories Comparison} and~\ref{fig: tracking errors Comparison}, the self-triggered coverage controller drove the $12$-agent MAS to a similar CVC as the continuous-time Lloyd algorithm. 
Moreover, the centroid tracking error of each agent, under the self-triggered coverage controller, may be bounded above by $1.26\times 10^{-12}$ for all $t\geq 149.5$.
While the difference in the final values of the centroid tracking errors between the benchmark and self-triggered coverage controller is negligible, the latter took about $1.87$ times longer than the former to minimize its centroid tracking errors.
In addition, if we treat the computational cost of a Voronoi cell, guaranteed Voronoi cell, and dual guaranteed Voronoi cell the same, then the total computational cost of implementing the self-triggered coverage controller for agent $p$ is approximately $28,000\cdot \zeta_p$---about $1.87$ times the total computational cost of agent $p$ for the benchmark case.
Recall, during an event time for agent $p$, agent $p$ must compute its Voronoi cell to determine its Voronoi cell centroid and update its controller.
Furthermore, agent $p$ may need to repeatedly compute/propagate the self trigger function $h_p(s)$ to determine its future event time, which requires the computation of guaranteed and dual guaranteed Voronoi cells.
These computations are nontrivial and may be thought of as having a similar cost to that of a Voronoi cell.
Hence, we lump the computational costs of a Voronoi cell, guaranteed Voronoi cell, and dual guaranteed Voronoi cell together.
To simplify the comparison, we neglect the nontrivial computational costs of the diameter of a dual guaranteed Voronoi cell and the distance between the centroid of a guaranteed Voronoi cell and its boundary, which are items needed to compute the self trigger function $h_p(s)$.
Although the self-triggered coverage controller nearly \textit{doubles} the computational cost of the benchmark per agent, Figure~\ref{fig: timer plots Self Trigger} shows that the inter-event times of each agent tend to increase as the MAS approaches a CVC under the self-triggered coverage controller.
In fact, nearly all agents enjoy inter-event times bounded below by $0.1$ for all $t\geq 50$, which implies that the self-triggered coverage controller requires no more than about $1/10$ of the communication and/or sensing used in the benchmark case for $t\geq 50$.
Hence, the self-triggered coverage controller more efficiently consumes communication and sensing resources at the expense of added computation relative to the benchmark.

As demonstrated by Figures~\ref{fig: MAS trajectories Comparison} and~\ref{fig: tracking errors Comparison}, the timer-based coverage controller drove the $12$-agent MAS to a different CVC than that of the benchmark.
Nevertheless, the centroid tracking error of each agent may be bounded above by $8.71\times 10^{-14}$ for all $t\geq 105$. 
One can then see that the difference in the final values of the centroid tracking errors between the benchmark and timer-based coverage controller is negligible, and the latter took about $1.31$ times longer than the former to minimize its centroid tracking errors.
Figure~\ref{fig: Voronoi Count Comparison} shows that each agent computed their Voronoi cell no more than $400$ times, which leads to a total computational cost that is approximately $0.03$ and $0.01$ times that of the benchmark and self-triggered coverage controller, respectively.
Moreover, since $T_1^p = 0.2$ for each $p\in\verts$, the timer-based coverage controller enabled each agent to use no more than $1/20$ of the communication and/or sensing used in the benchmark case for $t\geq 0$.

As shown by this brief comparison, the timer-based coverage controller offers comparable error regulation performance relative to the continuous-time Lloyd algorithm and self-triggered coverage controller.
An advantage of the timer-based coverage controller is that it can operate with far less computation, communication, and/or sensing when compared to the continuous-time Lloyd algorithm and self-triggered coverage controller.
In \fred{actuality}, the timer-based coverage controller does not require communication provided position information is available via measurement.
Note, while the self-triggered coverage controller requires each agent to continuously measure its position, the timer-based coverage controller only requires position measurements during event times.

\section{Conclusion}
This article delves into coverage control for static, \fred{compact}, and convex workspaces.
A hybrid extension of the continuous-time Lloyd algorithm is developed using timer-based control, where the reliance on continuous communication and/or sensing is supplanted with intermittent position measurements that may be harvested asynchronously between agents.
Our timer-based coverage controller improves upon existing event/self-triggered formulations by eliminating the need for Voronoi cell estimation, wireless communication, and increased communication rates as the MAS approaches a configuration corresponding to a centroidal Voronoi tessellation.
Zeno behavior is excluded \textit{a priori} by uniformly lowering bounding the distance between consecutive \fred{event} times for each agent with a positive user-defined parameter, which can be selected according to hardware specifications.
The closed-loop ensemble dynamics is modeled as a hybrid system whose maximal solutions are shown to be complete and, under the satisfaction of \fred{specific} sufficient conditions, converge to a set encoding the $\nu$-approximate coverage objective.

As future work, one could consider a thorough investigation of the Jacobian of $\cent(x)$ to derive an exponential stability result, which the Lloyd algorithm seems capable of achieving.
Further, one could consider an alternative formulation of the coverage control problem that is more amenable to standard Lyapunov analysis methods or contraction theory, switching workspace geometries and/or density functions to accommodate dynamic environments, and generalizations of coverage on manifolds.

\bibliographystyle{IEEEtran}
\bibliography{References}

\end{document}